\documentclass[aps,prd,superscriptaddress,twocolumn,nofootinbib]{revtex4-2}
\usepackage[utf8]{inputenc}
\usepackage{dsfont}
\usepackage{amsfonts}
\usepackage{amsmath}
\allowdisplaybreaks[4]        
\usepackage{amssymb}
\usepackage{euscript}     
\usepackage{braket}
\usepackage{starfont}
\usepackage{color,soul}         
\usepackage{tensor}        
\usepackage{amsthm}
\usepackage{graphicx}
\usepackage{slashed}
\usepackage{leftidx}
\usepackage{subfigure}
\usepackage{bbm}
\usepackage{enumitem}
\definecolor{outerspace}{rgb}{0.25, 0.29, 0.3}
\definecolor{scarlet}{rgb}{1.0, 0.13, 0.0}
\usepackage[header,title,page,titletoc]{appendix}  
\definecolor{princetonorange}{rgb}{1.0, 0.56, 0.0}
\definecolor{WildStrawberry}{rgb}{1.0, 0.26, 0.64}
\definecolor{rossocorsa}{rgb}{0.83, 0.0, 0.0}
\definecolor{navyblue}{rgb}{0.0, 0.0, 0.5}
\usepackage{float}
\usepackage[paper=letterpaper,margin=1in]{geometry}

\usepackage{mathtools}

\DeclareMathAlphabet{\pazocal}{OMS}{zplm}{m}{n}
\newcommand{\req}[1]{(\ref{#1})} 

\newcommand{\bea}{\begin{eqnarray}}

\newcommand{\eea}{\end{eqnarray}}
\newcommand{\ba}{\begin{eqnarray}}
\newcommand{\ea}{\end{eqnarray}}

\newcommand{\be}{\begin{equation}}
\newcommand{\ee}{\end{equation} }
\newcommand{\beqa}{\begin{eqnarray}}
\newcommand{\eeqa}{\end{eqnarray}}
\newcommand{\beqar}{\begin{eqnarray*}}
\newcommand{\eeqar}{\end{eqnarray*}}

\renewcommand{\req}[1]{(\ref{#1})}

\newcommand{\ie}{{\it i.e.,}\ }

\makeatletter
\newcommand{\dal}{\mathop{\mathpalette\dal@\relax}}
\newcommand{\dal@}[2]{%
  \begingroup
  \sbox\z@{$\m@th#1\square$}%
  \dimen0=\fontdimen8
    \ifx#1\displaystyle\textfont\else
    \ifx#1\textstyle\textfont\else
    \ifx#1\scriptstyle\scriptfont\else
    \scriptscriptfont\fi\fi\fi3
  \makebox[\wd\z@]{%
    \hbox to \ht\z@{%
      \vrule width \dimen0
      \kern-\dimen0
      \vbox to \ht\z@{
        \hrule height \dimen0 width \ht\z@
        \vss
        \hrule height 2\dimen0
      }%
      \kern-2.5\dimen0
      \vrule width 2.5\dimen0
    }%
  }%
  \endgroup
}

\makeatother

\usepackage{hyperref}
\hypersetup{
    colorlinks,
    citecolor=navyblue,
    filecolor=navyblue,
    linkcolor=navyblue,
    urlcolor=navyblue
}

\begin{document}

\title{Regular black holes from thin-shell collapse}
\author{Pablo Bueno}
\email{pablobueno@ub.edu}
\affiliation{Departament de F\'isica Qu\`antica i Astrof\'isica, Institut de Ci\`encies del Cosmos\\
 Universitat de Barcelona, Mart\'i i Franqu\`es 1, E-08028 Barcelona, Spain }

\author{Pablo A. Cano}
\email{pablo.cano@icc.ub.edu}
\affiliation{Departament de F\'isica Qu\`antica i Astrof\'isica, Institut de Ci\`encies del Cosmos\\
 Universitat de Barcelona, Mart\'i i Franqu\`es 1, E-08028 Barcelona, Spain }

\author{Robie A. Hennigar}
\email{robie.a.hennigar@durham.ac.uk}
\affiliation{Departament de F\'isica Qu\`antica i Astrof\'isica, Institut de Ci\`encies del Cosmos\\
 Universitat de Barcelona, Mart\'i i Franqu\`es 1, E-08028 Barcelona, Spain }
\affiliation{Centre for Particle Theory, Department of Mathematical Sciences, Durham University, Durham DH1 3LE, U.K.}

\author{\'Angel J. Murcia}
\email{angelmurcia@icc.ub.edu}
\affiliation{Departament de F\'isica Qu\`antica i Astrof\'isica, Institut de Ci\`encies del Cosmos\\
 Universitat de Barcelona, Mart\'i i Franqu\`es 1, E-08028 Barcelona, Spain }
\affiliation{INFN, Sezione di Padova, Via Francesco Marzolo 8, I-35131 Padova, Italy}


\begin{abstract}
We establish that regular black holes can form from gravitational collapse. Our model builds on a recent construction that realized regular black holes as exact solutions to purely gravitational theories that incorporate an infinite tower of higher curvature corrections in any dimension $D \ge 5$~\cite{Bueno:2024dgm}. We identify a two-dimensional Horndeski theory that captures the spherically symmetric dynamics of the theories in question and use this to prove a Birkhoff theorem and obtain the generalized Israel junction conditions. Armed with these tools, we consider the collapse of thin shells of pressureless matter, showing that this leads generically to the formation of regular black holes. The interior dynamics we uncover is intricate, consisting of shell bounces and white hole explosions into a new universe. The result is that regular black holes are the unique spherically symmetric solutions of the corresponding theories and also the endpoint of gravitational collapse of matter. Along the way, we establish evidence for a solution-independent upper bound on the curvature, suggestive of Markov’s limiting curvature hypothesis.

\end{abstract}
\maketitle
\section{Introduction}

The Penrose-Hawking singularity theorems were a triumph of the early days of mathematical relativity and remain a cornerstone of modern gravitational theory~\cite{Penrose:1964wq, Hawking:1973uf}. The theorems rigorously establish that singularities are the generic outcome of gravitational collapse in General Relativity, a discovery which earned Penrose his share of the 2020 Nobel Prize in Physics. However, this triumph also revealed a foundational problem. Our theories of Nature breakdown at singularities, and if singularities are real, it would mean an unassailable indeterminacy in our description of reality. It is therefore widely expected that singularities will be resolved by quantum gravitational effects. There are good reasons for this expectation --- the singularity theorems assume (among other things) classical energy conditions, which can be violated by quantum matter. 


In the absence of a known mechanism for singularity resolution, many authors have considered its \textit{phenomenological} implications. Perhaps the paradigmatic example of this is the programme of research on regular black holes. Dating to the early work of the 1960s~\cite{Sakharov:1966aja,1968qtr..conf...87B}, authors have considered models of black holes with singularity-free (regular) cores and studied their geometrical properties. Importantly, as these metrics are postulated on heuristic grounds, they do not \textit{a priori} solve the equations of motion of any known theory. Nevertheless, it has been possible to learn a variety of general features that apply to broad classes of regular black hole metrics. For example, it is often --- but not always --- necessary to violate energy conditions in the construction of these metrics~\cite{Mars:1996khm,Borde:1996df}. Moreover, regular black holes provide a setting in which a self-consistent picture of the black hole evaporation process can be studied, at least in principle~\cite{Hayward:2005gi, Frolov:2014jva}. On the other hand, all known examples of regular black hole metrics possess an inner Cauchy horizon, suggesting potential instabilities~\cite{DiFilippo:2022qkl, Carballo-Rubio:2022kad}. There is a vast  literature on the phenomenology of regular black holes, which has recently received renewed attention~\cite{Carballo-Rubio:2018pmi, Carballo-Rubio:2018jzw, Carballo-Rubio:2019nel, Carballo-Rubio:2019fnb,Franzin:2022wai,Ghosh:2022gka, Vagnozzi:2022moj,Pedrotti:2024znu,Calza:2024xdh,Calza:2024fzo, Davies:2024ysj}.

While phenomenological considerations can reveal much, they cannot tell the complete story. To fully assess the viability of regular black holes, at some point it becomes necessary to find these metrics as solutions of bona fide theories and to prove that they are the outcome of gravitational collapse. Over the years, there have been various approaches to finding regular black holes as \textit{solutions}, \textit{e.g.},~\cite{Frolov:1989pf,Barrabes:1995nk,Easson:2002tg, Nicolini:2005vd,Olmo:2012nx,Balakin:2015gpq,Bazeia:2015uia,Chamseddine:2016ktu,Bambi:2016xme, Easson:2017pfe, Bejarano:2017fgz,Colleaux:2017ibe,Cano:2018aod, Colleaux:2019ckh,Cano:2020qhy,Cano:2020ezi,Guerrero:2020uhn,Bueno:2021krl,Brandenberger:2021jqs,Olmo:2022cui,Biasi:2022ktq,Junior:2024xmm,Ovalle:2024wtv,Alencar:2024yvh,Bronnikov:2024izh,Bolokhov:2024sdy,Skvortsova:2024wly}.\footnote{Another avenue to motivate the existence of regular black holes is to include quantum effects, such as those coming from the semi-classical Einstein equations or from the renormalization of Newton's constant. See \textit{e.g.} \cite{Bonanno:2000ep,Greenwood:2008ht,Saini:2014qpa,Kawai:2017txu}.} The approach that has garnered the most attention has been that of nonlinear electrodynamics~\cite{Ayon-Beato:1998hmi,Bronnikov:2000vy,Ayon-Beato:2000mjt,Bronnikov:2000yz,Ayon-Beato:2004ywd,Dymnikova:2004zc,Berej:2006cc,Balart:2014jia,Bronnikov:2017sgg,Junior:2023ixh,Murk:2024nod,Li:2024rbw,Zhang:2024ljd}. In fact, as nicely detailed in~\cite{Fan:2016rih}, this can often be reduced to a procedural exercise: given a regular black hole metric, a theory of nonlinear electrodynamics can be constructed that admits it as a solution. 

However, this perspective is not fully satisfactory. There are a number of important and pervasive issues associated with regular black holes in nonlinear electrodynamics. For example, the theories required often do not recover Maxwell electrodynamics in the weak field limit, thus postulating new and exotic matter for singularity resolution. Moreover, this matter is \textit{required} for singularity resolution, with the fully singular Schwarzschild metric being recovered when the charged matter is switched off.   Another issue is that the regular black holes, while indeed solutions of these theories, are not the \textit{general solutions}. Instead, they require fine tuning between integration and coupling constants~\cite{Fan:2016rih}. Therefore, the most general solutions of these theories remain singular, even in spherical symmetry. 

A different approach to the problem, which is capable of resolving some of the above issues, is to consider higher derivative corrections featuring nonminimal couplings to matter. In fact, two of us have previously shown that regular, electrically charged black holes may be found in four dimensions as solutions to certain higher-derivative extensions of Einstein-Maxwell theory \cite{Cano:2020qhy,Cano:2020ezi}. Remarkably, the mass and the charge can vary independently in this setup, without the need for fine tuning of coupling and integration constants. A similar approach applies in three dimensions \cite{Bueno:2021krl} and likely also in $D \ge 5$~\cite{Cano:2022ord}. Nevertheless, all of these models continue to suffer from the issue of \textit{requiring} charge for singularity resolution, recovering the Schwarzschild singularity when the charged matter is switched off. To address this problem, a purely gravitational mechanism for singularity resolution seems to be required.


Recently, three of us have put forth a model in which regular black holes are obtained as exact solutions of a purely gravitational theory~\cite{Bueno:2024dgm}. More specifically, it was shown that when the Einstein-Hilbert action is supplemented by an infinite tower of higher curvature corrections, then the Schwarzschild singularity is resolved \textit{generically} in any $D \ge 5$. The approach makes use of a class of gravitational theories known as \textit{quasi-topological gravities}~\cite{Oliva:2010eb,Quasi,Dehghani:2011vu,Ahmed:2017jod,Cisterna:2017umf,Bueno:2019ycr, Bueno:2022res, Moreno:2023rfl}. These theories --- as we review in detail below --- are constructed to have particularly desirable properties in spherical symmetry. They exist at any order in curvature for any $D \ge 5$~\cite{Bueno:2019ycr, Moreno:2023rfl} and, notably, for a static and spherically symmetric ansatz, the equations of motion reduce to a single algebraic equation that determines the metric function. The key observation of~\cite{Bueno:2024dgm} was that \textit{precisely} when an infinite number of quasi-topological densities are included in the action, the black hole singularity is resolved. This is achieved without fine tuning, requiring only very mild assumptions on the coupling constants of the different terms pertaining to their sign and relative growth. 


One may worry that selecting a particular class of theories makes the results finely tuned in the \textit{space of theories}, even if not in the space of solutions to those theories. However, quasi-topological gravities are sufficiently general to provide a basis for the gravitational effective action in vacuum~\cite{Bueno:2019ltp, Bueno:2024dgm}. Therefore, in the regime of effective field theory, quasi-topological gravities capture the \textit{most general} corrections to General Relativity for any $D \ge 5$. While the resolution of the singularity requires going beyond the regime of effective field theory (because the full tower of corrections is being resummed), one is free to instead treat the couplings perturbatively, in which case regular black hole metrics can be matched to \textit{any desired accuracy} in the perturbative expansion. All together, this provides strong support for the generality of the results. 

In this paper we tackle a different question --- the formation of regular black holes from gravitational collapse. Until now, the study of the dynamical collapse of matter into a regular black hole solution of the theory controlling such dynamics had remained out of reach. Here, we accomplish this goal by extending and improving upon the results of~\cite{Bueno:2024dgm} in several ways.



A key tool in our analysis is the identification of a two-dimensional Horndeski theory that captures the spherically symmetric sector of quasi-topological gravities. This two-dimensional theory is obtained directly, by dimensionally reducing the higher-dimensional theories on the sphere. Armed with this, we are able to present explicit, fully resummed and covariant two-dimensional actions that describe the corresponding fully resummed spherical compactifications of the higher-dimensional theories. Using this, we prove that a Birkhoff theorem holds at all orders in quasi-topological gravity, so that the static and spherically symmetric regular black hole solutions considered in~\cite{Bueno:2024dgm} are in fact the \textit{unique} spherically symmetric solutions of the corresponding theories. We then obtain the generalization of the Israel junction conditions for the theories at hand and study the problem of spherically symmetric thin shell collapse. In General Relativity, the collapse of a thin shell results in the formation of the Schwarzschild metric in finite time, including its singular interior. In stark contrast to this, in our setup we find that regular black holes generically result from the collapse of matter.

The dynamics of the thin shells is intricate, but exhibits general patterns. We observe that a shell, beginning from rest, collapses leaving behind one of the regular black hole solutions of~\cite{Bueno:2024dgm} in the exterior. In the interior, the shells continue to collapse before ultimately reaching a turning point. As we establish using a master formula for the shell dynamics, this occurs regardless of the particular model considered. Then, a bounce occurs, with the shell rebounding and emerging into a new universe through a white hole explosion. The situation is very similar to that considered in various quantum gravity inspired models~\cite{Hajicek:2001yd,Haggard:2014rza, Husain:2021ojz, Han:2023wxg}, here emerging naturally from the dynamics of the underlying theory.  

The result of all this is an explicit model in which regular black holes are the unique solutions to the corresponding equations of motion, and arise as the product of gravitational collapse, generically, in any dimension $D \ge 5$. While the mechanism identified here may not be the one ultimately responsible for singularity resolution in Nature, what is clear is that it provides the first comprehensive approach to the programme of regular black holes. Without further ado, let us begin the analysis with a detailed description of quasi-topological theories.

\section{Quasi-topological Gravities}
Let us consider a general $D$-dimensional theory of gravity built from contractions of the Riemann tensor and the metric,
\begin{equation}\label{QTaction}
I= \frac{1}{16\pi G_{\rm N} }\int \mathrm{d}^Dx \sqrt{|g|}\mathcal{L}(g^{ab},R_{cdef})\, ,
\end{equation}
where $G_{\rm N}$ is Newton's constant.
 In the absence of matter, the equations of motion read
\begin{align}\label{eq:Eab}
\mathcal{E}_{ab} 
=P_a{}^{cde}R_{bcde}-\frac{1}{2}g_{ab}\mathcal{L}-2\nabla^c\nabla^d P_{acdb}=0\, ,  
\end{align}
where we defined
\begin{equation}
P^{abcd}\equiv \left[ \frac{\partial \mathcal{L}}{\partial R_{abcd}}\right]\, ,
\end{equation}
which inherits the symmetries of the Riemann tensor. Naturally, the above equations are of fourth order in derivatives of the metric, except in the case in which the third term is absent, namely, when
\be  
\nabla^d P_{acdb}=0 \quad \Leftrightarrow \quad \text{Lovelock gravity}
\ee
As indicated above, this can be seen as the defining property of Lovelock gravities \cite{Lovelock1,Lovelock2,Padmanabhan:2013xyr}. For those, the equations of motion are second-order for general spacetimes. They exist and have non-trivial dynamics for curvature orders $n \leq \lfloor (D-1)/2 \rfloor$ and they are topological in the critical dimension $n=D/2$. Hence, for any spacetime dimension $D$ there is always a finite number of non-trivial Lovelock densities. In other words, reaching arbitrarily high-order non-trivial Lovelock densities would require $D\rightarrow \infty$.

In this paper we shall consider a more general class of higher-curvature gravities which, in any number of dimensions $D\geq 5$, contains non-trivial densities of arbitrarily high order. In order to define it, let us consider a general static and spherically symmetric (SSS) ansatz,
\begin{equation}\label{Nf}
\mathrm{d}s^2_{\rm SSS}=-N(r)f(r)\mathrm{d}t^2+\frac{\mathrm{d}r^2}{f(r)}+r^2\mathrm{d}\Omega_{(D-2)}^2\, .
\end{equation}
We say that our higher-curvature gravity is of the quasi-topological (QT) class whenever 
\be \label{SSSs}
\left.\nabla^d P_{acdb}\right|_{ f}=0 \quad \Leftrightarrow \quad \text{QT gravity}
\ee
where $\vert_{ f}$ stands for evaluation on the SSS ansatz \eqref{Nf} with $N(r)=1$. This makes it possible to solve the equations and find black hole solutions analytically. Explicit forms for QT Lagrangians of arbitrary order can be found in \cite{Bueno:2019ycr,Moreno:2023rfl}. 

\subsection{QT gravities with a Birkhoff theorem}
QT gravities have an important level of degeneracy, in the sense that, at a given curvature order, one finds different QT Lagrangians that yield identical equations for SSS metrics. All these Lagrangians are equivalent from the point of view of SSS solutions, but their degeneracy is generically broken if one considers more general classes of solutions. This allows one to identify subsets of QT Lagrangians that satisfy further constraints \cite{Bueno:2018uoy,Arciniega:2018fxj,Arciniega:2018tnn,Moreno:2023arp}.  In this paper, we consider a more restricted class of QT gravities, defined by the condition that they possess second-order equations of motion on general spherically symmetric (SS) metrics. Since we do not impose staticity, this condition is stronger than the usual definition of QT gravities based on SSS metrics with $N(r)=1$. These theories will be especially well-suited to study spherical collapse; having second-order equations of motion, one can analyze time evolution avoiding the pathologies associated with higher-order time derivatives. On the other hand, all these theories satisfy a Birkhoff theorem, as we show below, implying the uniqueness of spherical black hole solutions. The first examples of theories of this kind were found by \cite{Oliva:2010eb,Oliva:2011xu}, and here we extend those results to arbitrary high orders in the curvature in any dimension $D\ge 5$. We will refer to this family of theories by ``Birkhoff QT gravities'' or simply ``Birkhoff gravities''. 

At the lowest orders in the curvature, one can find instances of these theories by a brute force selection of Lagrangians. The idea is 
\begin{enumerate}[leftmargin=*]
\item Write down a general (or at least, general enough) order-$n$ Lagrangian density, $\mathcal{Z}_{n}$, consisting of a sum of curvature invariants with free couplings.
\item Compute the equations of motion \req{eq:Eab} of this Lagrangian and evaluate them on a general SS (time-dependent) ansatz.
\item Fix the couplings of the Lagrangian so that the terms with more than two derivatives in the equations of motion cancel out. 
\end{enumerate}
Remarkably, this process leads to a non-empty set of theories beyond the Lovelock family, in dimensions $D\ge 5$. The following expressions are explicit instances of such theories up to order $n=5$ and arbitrary dimension: 
\begin{widetext}
\begin{subequations}\label{Znexplicit}
\begin{align}
 \mathcal{Z}_{(1)}&=R\,, \\
 \mathcal{Z}_{(2)}&=\frac{1}{(D-2)} \left [\frac{W_{abcd} W^{abcd}}{D-3} -\frac{4 Z_{ab}Z^{ab}}{D-2}\right]+\frac{\mathcal{Z}_{(1)}^2}{D(D-1)}\,, \\
\nonumber \mathcal{Z}_{(3)}&=\frac{24}{(D-2)(D-3)} \left[\frac{ W\indices{_a_c^b^d} Z^a_b Z^c_d}{(D-2)^2}-
   \frac{   W_{a c d e}W^{bcde}Z^a_b}{(D-2) (D-4)}+\frac{2(D-3)
   Z^a_b Z^b_cZ_a^c}{3(D-2)^3} +\frac{(2 D-3) W\indices{^a^b_c_d}W\indices{^c^d_e_f}W\indices{^e^f_a_b}}{12 (D ((D-9)
   D+26)-22)} \right]\\&+ \frac{3\mathcal{Z}_{(1)}\mathcal{Z}_{(2)}}{D(D-1)}-\frac{2 \mathcal{Z}_{(1)}^3}{D^2(D-1)^2} \,, \\
\nonumber
\mathcal{Z}_{(4)}&=\frac{96}{(D-2)^2(D-3)} \left[\frac{(D-1)\left ( W_{abcd} W^{abcd} \right)^2}{8D(D-2)^2(D-3)}-\frac{(2D-3) Z_e^f Z^e_f W_{abcd} W^{abcd}}{4(D-1)(D-2)^2}-
\frac{2 W_{acbd} W^{c efg} W^d{}_{efg} Z^{ab} }{D(D-3)(D-4)}\right. \\ \nonumber & -\frac{4Z_{a c} Z_{d e} W^{bdce} Z^{a}_b}{(D-2)^2(D-4)} \left. +\frac{(D^2-3D+3) \left (Z_a^b Z_b^a\right )^2}{D(D-1)(D-2)^3}-\frac{Z_a^b Z_b^c Z_c^d Z_d^a}{(D-2)^3}+\frac{(2D-1)W_{abcd} W^{aecf} Z^{bd} Z_{ef}}{D(D-2)(D-3)}\right]\\&+\frac{4\mathcal{Z}_{(1)}\mathcal{Z}_{(3)}-3 \mathcal{Z}_{(2)}^2}{D(D-1)}\,,\\ \nonumber
\mathcal{Z}_{(5)}&=\frac{960 (D-1)}{(D-2)^4(D-3)^2} \left[ \frac{(D-2)W_{ghij} W^{ghij}W\indices{_a_b^c^d}W\indices{_c_d^e^f}W\indices{_e_f^a^b} }{40D(D^3-9 D^2+26D-22)}+\frac{4(D-3) Z_a^b Z_b^c Z_c^d Z_d^e Z_e^a}{5(D-1)(D-2)^2(D-4)}\right. \\ \nonumber & -\frac{(3D
-1)W^{ghij} W_{ghij}  W_{a c d e}W^{bcde} Z^a_b}{10D(D-1)^2(D-4)}-\frac{4(D-3)(D^2-2D+2)Z_a^b Z_b^a Z_c^d Z_d^e Z_e^c}{5D(D-1)^2(D-2)^2(D-4)} \\ \nonumber & -\frac{(D-3)(3D-1)(D^2+2D-4)W^{ghij} W_{ghij} Z_c^d Z_d^e Z_e^c}{10D(D-1)^2(D+1)(D-2)^2(D-4)}+\frac{(5D^2-7D+6)Z_g^h Z_h^g W_{abcd} Z^{ac} Z^{bd}}{10D(D-1)^2(D-2)}\\ \nonumber & +\frac{(D-2)(D-3)(15D^5-148 D^4+527 D^3-800 D^2+472D-88)W\indices{_a_b^c^d}W\indices{_c_d^e^f}W\indices{_e_f^a^b} Z_{g}^h Z_h^g}{40D(D-1)^2(D-4)(D^5-15D^4+91 D^3-277 D^2+418D-242)}\\ \nonumber &- \frac{2(3D-1)Z^{ab} W_{acbd} Z^{ef}  W\indices{_e^c_f^g} Z^d_g}{D(D^2-1)(D-4)}-\frac{Z_{a}^b Z_{b}^{c} Z_{cd} Z_{ef} W^{eafd}}{(D-1)(D-2)} +\frac{(D-3)W_{a c d e}W^{bcde} Z^a_b Z_f^g Z_g^f}{5D(D-1)^2(D-4)}\\ \nonumber &\left. -\frac{(D-2)(D-3)(3D-2) Z^a_b Z^b_c W_{daef} W^{efgh} W_{gh}{}^{dc}}{4(D-1)^2(D-4)(D^2-6D+11)}+\frac{W_{ghij}W^{ghij} Z^{ac}Z^{bd}W_{abcd}}{20D(D-1)^2}\right]\\&+\frac{5\mathcal{Z}_{(1)}\mathcal{Z}_{(4)}-2\mathcal{Z}_{(2)}\mathcal{Z}_{(3)}}{D(D-1)}+\frac{6 \mathcal{Z}_{(1)}\mathcal{Z}_{(2)}^2-8 \mathcal{Z}_{(1)}^2\mathcal{Z}_{(3)}}{D^2(D-1)^2}\,.
\end{align}
\end{subequations}
\end{widetext}
Here  $W_{abcd}$ is the Weyl curvature tensor and 
\begin{equation}
Z_{ab}\equiv R_{ab}-\frac{1}{D}g_{ab}R\, ,
\end{equation}
is the traceless part of the Ricci tensor. Naturally, the density with $n=1$ is the Einstein-Hilbert term, while the quadratic density $n=2$ is the Gauss-Bonnet invariant (expressed in an unconventional way). The densities with $n\ge 3$ no longer belong to the Lovelock class, and furthermore, they are not unique (\textit{e.g.}, one can modify the Lagrangian by adding terms that vanish on spherically symmetric metrics).  
However, in order to analyze the properties of these theories, and to extend them to higher orders, we need a more refined way of identifying them. We address this next by studying the reduced two-dimensional action of the Birkhoff gravities.

 \subsection{Effective two-dimensional action}\label{subi}
Let us consider an ansatz for a general $D$-dimensional spherically symmetric metric, 
\begin{equation}\label{sphericmetric}
\mathrm{d}s^2=\gamma_{\mu\nu}\mathrm{d} x^{\mu}\mathrm{d}x^{\nu}+\varphi(x)^2 \mathrm{d}\Omega^2_{D-2}\, ,
\end{equation}
where $\gamma_{\mu\nu}$ is a 2-dimensional metric --- we will use Greek indices to refer to components of this two-dimensional Lorentzian submanifold ---, $\mathrm{d}\Omega^2_{D-2}$ is the metric of the $(D-2)$-sphere, and $\varphi(x)$ plays the role of a covariant radial coordinate. The evaluation of a higher-dimensional gravitational action \req{QTaction} on this ansatz yields a two-dimensional ``dilaton gravity'' model for the metric $\gamma_{\mu\nu}$ and the scalar $\varphi$~\cite{Frolov:1998wf, Grumiller:2002nm}. The key realization is the following: if the higher-dimensional theory possesses second-order equations for spherically symmetric metrics, then its reduced action must be a two-dimensional Horndeski theory, since Horndeski Lagrangians capture the most general scalar-tensor theory with second-order equations of motion \cite{Horndeski:1974wa}. Let us explicitly verify that this phenomenon holds true for the five Birkhoff-QTs we presented in \eqref{Znexplicit}.

To this aim, we first note that the $D$-dimensional Weyl curvature tensor $W_{abcd}$,  traceless Ricci curvature tensor $Z_{ab}$ and the Ricci scalar $R^{(D)}$ (we momentarily include the $D$ superscript to distinguish it from the Ricci scalar of the two-dimensional metric $\gamma_{\mu \nu}$, suppressing the $D$ superscript whenever no confusion may arise) on top of \eqref{sphericmetric} take the following form:
\begin{align}
\nonumber
W_{ab}{}^{cd}&=\Omega \left[\frac{(D-2)(D-3)}{2} \gamma_{[a}{}^{c} \gamma_{b]}{}^{d}+\sigma_{[a}{}^{c} \sigma_{b]}{}^{d}\right.  \\ \label{eq:weyldec}& \left. -(D-3) \gamma_{[a}{}^{[c} \sigma_{b]}{}^{d]} \right] \,, \\
\label{eq:riccidec}
Z_{ab}&=\delta_{a}^\mu \delta_{b}^\nu \mathcal{S}_{\mu \nu}+\Theta \, \sigma_{ab}\,,  \\
\label{eq:scaldec}
R^{(D)}&=R-(D-2) \left[ \frac{2\Box \varphi}{\varphi}-(D-3) \psi \right] \,,
\end{align}
where $\gamma_{ab}=\delta_{a}^\mu \delta_{b}^\nu \gamma_{\mu \nu}$, $\sigma_{ab}=g_{ab}-\gamma_{ab}$, $R$ is the Ricci scalar of $\gamma_{\mu \nu}$, $\Box \varphi$ is the Laplacian of $\varphi$ associated with $\gamma_{\mu \nu}$ and:
\begin{align}
\label{eq:psiXdef}
\psi&=\frac{1-X}{\varphi^2}\, ,\quad X=\nabla_{\mu}\varphi\nabla^{\mu}\varphi\, , \\
\Omega&=\frac{2(2\varphi^2 \psi+2 \varphi \Box \varphi +\varphi^2 R)}{(D-1)(D-2) \varphi^2}\,,\\
\Theta&=\frac{2(D-3)\varphi^2 \psi+(D-4) \varphi \Box \varphi-\varphi^2 R}{D \varphi^2}\,,\\
\nonumber
\mathcal{S}_{\mu \nu}&= \Xi \gamma_{\mu \nu}  -(D-2)\frac{\nabla_\mu \nabla_\nu \varphi}{\varphi}  \,, \\
\Xi&= \frac{(D-2)}{D}\left (\frac{R}{2}+\frac{2 \Box \varphi}{\varphi}-(D-3) \psi \right)\,.
\end{align}
Note that $\mathcal{S}_{\mu \nu}$ is a symmetric two-dimensional tensor transverse to the spherical sections 
and $\nabla_{\mu}$ is the covariant derivative associated with $\gamma_{\mu\nu}$. The following properties hold:
\begin{align}
&\gamma_{ab} \gamma^{bc}=\gamma_{a}{}^{c}\,, \quad \gamma_{ab} \sigma^{bc}=0\,, \quad \sigma_{ab} \sigma^{bc}=\sigma_{a}{}^{c}\,,\\
& \gamma_{a}{}^{a}=2\,, \quad \mathcal{S}_{\mu}{}^{\mu}=-\sigma_{a}{}^{a} \Theta\,, \quad  \sigma_{a}{}^{a}=D-2 \,.
\end{align}
Using the decomposition given at \eqref{eq:weyldec}, \eqref{eq:riccidec} and \eqref{eq:scaldec}, we conclude that all $D$-dimensional curvature invariants evaluated on \eqref{sphericmetric} will be entirely expressed in terms of $\Omega$, $\Theta$ and contractions of $\mathcal{S}_{\mu \nu}$. However, we can specialize even more, since the trace of $\mathcal{S}_{\mu \nu}$ is given in terms of $\Theta$  and contractions of more than two $\mathcal{S}_{\mu \nu}$ tensors will be related to products of lower order ones through the use of \emph{Schouten identities} --- \ie expressions derived from the antisymmetrizations of more than two indices, which are identically zero for two-dimensional tensors. Therefore, every $D$-dimensional curvature invariant on top of \eqref{sphericmetric} may be purely written in terms of $\Omega$, $\Theta$ and $\mathcal{S}_{\mu \nu} \mathcal{S}^{\mu \nu}$, which takes the form:
\begin{align}
\nonumber
\mathcal{S}_{\mu \nu} \mathcal{S}^{\mu \nu}&=2 \Xi^2-\frac{2 (D-2) \Xi \Box \varphi}{\varphi} \\&+\frac{
(D-2)^2 \nabla_\mu \nabla_\nu \varphi \nabla^\mu \nabla^\nu \varphi }{\varphi^2} \,.
\end{align}
After these preliminaries, we are in position to write the various $D$-dimensional curvature invariants forming the five Birkhoff-QTs in terms of $\Omega$, $\Theta$ and $\mathcal{S}_{\mu \nu} \mathcal{S}^{\mu \nu}$. We already did this with the $D$-dimensional Ricci scalar in \eqref{eq:scaldec}, while the explicit expressions for the remaining quadratic, cubic, quartic and quintic densities appearing in \eqref{Znexplicit} are collected in appendix \ref{app2}.
Using those results, the evaluation of the five Birkhoff-QT densities $\mathcal{Z}_{n}$ \req{Znexplicit} on the metric \req{sphericmetric} can be dramatically simplified down to the following compact expression:
\begin{align}\label{hornZ}
\mathcal{Z}^{(\rm 2d)}_{n}=&+H_{2}^{(n)}(\varphi, X)-\Box\varphi H^{(n)}_{3}(\varphi, X)\\\notag&+H^{(n)}_{4}(\varphi, X)R  \\\notag
&-2H^{(n)}_{4,X}(\varphi, X)\left[(\Box\varphi)^2-\nabla_{\mu}\nabla_{\nu}\varphi\nabla^{\mu}\nabla^{\nu}\varphi\right]\, ,
\end{align}
where 
\begin{equation}\label{eq:HnHorn}
\begin{aligned}
H^{(n)}_{2}&=(D-2n)(D-2n-1)\psi^{n}\, ,\\
H^{(n)}_{3}&=2n(D-2n)\varphi^{-1}\psi^{n-1}\, ,\\
H^{(n)}_{4}&=n\psi^{n-1}\, ,
\end{aligned}
\end{equation}
where we used the short-hand notation $H_{p,X}\equiv \partial_{X}H_{p}$. For arbitrary functions $H_{p}(\varphi, X)$, the expression \req{hornZ} is nothing but the general form of the Horndeski Lagrangian in two dimensions. Thus, the reduced Birkhoff-QT densities yield a particular family of Horndeski theories, given by the functions \req{eq:HnHorn} at each order $n=1,2,3,4,5$. 
One could wonder if there are additional families of densities that generate other $H^{(n)}_{p}(\varphi, X)$ functions different from \req{eq:HnHorn}. We have found no evidence of this, as all the theories that we found fit into the pattern \req{eq:HnHorn}. Therefore, even though one can obtain multiple Birkhoff-QT densities at a given order, all of them yield the same equations of motion for spherically symmetric metrics, and it is enough to just consider one density at each order. 

With the aid of the five densities $\mathcal{Z}_{(n)}$ with $n=1,2,3,4,5$ presented in \req{Znexplicit}, it is possible to construct $D$-dimensional higher-curvature theories of the Birkhoff-QT class at arbitrary curvature order. This may be done by applying the following argumentation:
\begin{enumerate}[leftmargin=*]
\item Extend the definition of the two-dimensional Horndeski theories \eqref{hornZ} for arbitrary $n \geq 1$. For $n >5$, it is not clear at this moment if it might correspond to the dimensional reduction of a certain $n$-th order Birkhoff-QT theory on \eqref{sphericmetric}.
\item Now, we observe that the two-dimensional Hordneski theories \eqref{hornZ} for any $n \geq 1$ satisfy the following recursive relation:
 \begin{align}\notag
\mathcal{Z}_{n+5}^{(\rm 2d)}&=\frac{3(n+3)\mathcal{Z}_{1}^{(\rm 2d)}\mathcal{Z}_{n+4}^{(\rm 2d)}}{D(D-1)(n+1)}-\frac{3(n+4)\mathcal{Z}_{2}^{(\rm 2d)}\mathcal{Z}_{n+3}^{(\rm 2d)}}{D(D-1)n}\\ \label{zrec2d}&+\frac{(n+3)(n+4)\mathcal{Z}_{3}^{(\rm 2d)}\mathcal{Z}_{n+2}^{(\rm 2d)}}{D(D-1)n(n+1)}\, .
\end{align}
\item Starting now from the five densities $\mathcal{Z}_{(m)}$ with $m=1,2,3,4,5$ obtained in \eqref{Znexplicit}, consider the $D$-dimensional higher-curvature theory of order $n$ constructed inductively by formally substituting $\mathcal{Z}_{n}^{(\rm 2d)}$ in \eqref{zrec2d} by $\mathcal{Z}_{n}$:
 \begin{align}\notag
\mathcal{Z}_{n+5}&=\frac{3(n+3)\mathcal{Z}_{1}\mathcal{Z}_{n+4}}{D(D-1)(n+1)}-\frac{3(n+4)\mathcal{Z}_{2}\mathcal{Z}_{n+3}}{D(D-1)n}\\ \label{zrec}&+\frac{(n+3)(n+4)\mathcal{Z}_{3}\mathcal{Z}_{n+2}}{D(D-1)n(n+1)}\, .
\end{align}
Given the first five densities $\mathcal{Z}_{(m)}$ with $m=1,2,3,4,5$, it is clear than one obtains definite, explicit and unique expressions for $\mathcal{Z}_{(6)}$, $\mathcal{Z}_{(7)}...$, arriving to an arbitrary curvature order $n$ if desired. We have not shown yet whether these latter theories are Birkhoff-QTs.
\item Note that the first five densities $\mathcal{Z}_{(n)}$ presented in \eqref{Znexplicit} reduce to the Horndeski theory \eqref{hornZ} on top of \eqref{sphericmetric}, with the latter satisfying \eqref{zrec2d} for arbitrary curvature order $n$. Therefore, those higher-curvature gravities obtained from \eqref{zrec} at any curvature order $n$ will precisely boil down to \eqref{hornZ} when evaluated on \eqref{sphericmetric}. Consequently, they will have second-order equations of motion for any configuration \eqref{sphericmetric} and will belong to the Birkhoff-QT class. 
\end{enumerate}

In sum, we conclude that:
\begin{itemize}[leftmargin=*]
\item There exist Birkhoff-QT theories at any curvature order $n \geq 1$ and dimension $D \geq 5$. 
\item These may be found via \eqref{zrec}.
\item Using \eqref{zrec}, the subsequent theories take the form \req{hornZ} when evaluated on a spherically symmetric ansatz \req{sphericmetric}. 
\end{itemize}
Interestingly, the recursive formula \req{hornZ} is identical to the one found in \cite{Bueno:2019ycr} for ordinary QT gravities, but we have found that if one uses Birkhoff-QT densities as a seed, then the recursive formula also generates Birkhoff-QT gravities.

Let us then consider the full Birkhoff-QT action with an infinite number of terms, 
\begin{equation}\label{QTaction}
S= \frac{1}{16\pi G_{\rm N} }\int \mathrm{d}^Dx \sqrt{|g|}\left[R+\sum_{n=2}^{\infty}\alpha_{n}\mathcal{Z}_{n}\right]\, ,
\end{equation}
where we have set $\alpha_{1}=1$ so that the ($D$-dimensional) Einstein-Hilbert term is canonically normalized. 
By the previous argumentation, the reduction of this action on the ansatz \req{sphericmetric} yields the following two-dimensional theory 
\begin{equation}\label{eq:2daction}
S_{\rm 2d}=\frac{(D-2)\Omega_{D-2}}{16\pi G_{\rm N}}\int \mathrm{d}^{2}x\sqrt{|\gamma|} \mathcal{L}_{\rm 2d}(\gamma_{\mu\nu},\varphi)\, ,
\end{equation}
where  $\Omega_{(D-2)}=2\pi^{(D-1)/2}/\Gamma\left[\tfrac{D-1}{2}\right]$ is the volume of the $(D-2)$-sphere, 
\begin{align}\notag 
&\mathcal{L}_{\rm 2d}=G_{2}(\varphi, X)-\Box\varphi G_{3}(\varphi, X)+G_{4}(\varphi, X)R\\  \label{horn}&-2G_{4,X}(\varphi, X)\left[(\Box\varphi)^2-\nabla_{\mu}\nabla_{\nu}\varphi\nabla^{\mu}\nabla^{\nu}\varphi\right]\, ,
\end{align}
and the $G_{p}$ functions read
\begin{align}
G_{p}(\varphi, X)=&\frac{1}{D-2}\varphi^{D-2}\sum_{n=1}^{\infty}\alpha_{n}H^{(n)}_{p}\, .
\label{eq:defgp}
\end{align}
It is useful to rewrite these functions in terms of the following  ``characteristic polynomial'' $h(\psi)$, 
 \be \label{eom_psi}
h(\psi)\equiv \psi + \sum_{n=2}^{\infty}\alpha_n \frac{D-2n}{D-2}  \psi^n\, ,
\ee
which encapsulates many features of QT gravity solutions, as we shall see. In terms of this function, we obtain the expressions

\begin{align}
G_{2}(\varphi, X)&=\varphi^{D-2}\left[(D-1)h(\psi)-2\psi  h'(\psi)\right]\, ,\\\label{G3form}
G_{3}(\varphi, X)&=2\varphi^{D-3}h'(\psi)\, ,\\
\label{eq:G4}
G_{4}(\varphi, X)&=-\frac{1}{2}\varphi^{D-2}\psi^{(D-2)/2}\int \mathrm{d}\psi \psi^{-D/2}h'(\psi)\, ,
\end{align}
where $h'(\psi)\equiv dh(\psi)/d\psi$, and one should plug in the value of $\psi$ \req{eq:psiXdef} once the derivatives and integral have been computed. In particular, in \eqref{eq:G4} one should choose the primitive ensuring that $G_4$ matches precisely the definition in \eqref{eq:defgp}. In another vein, a crucial aspect of these three functions is that, since all of them are determined by $h(\psi)$, they are not independent and satisfy several relations. In fact, we get 
\begin{equation}
G_{4,\varphi}=\frac{1}{2}G_{3}\, ,\quad G_{2,X}=-\frac{1}{2}G_{3,\varphi}\, .
\end{equation}
These identities will be relevant for the simplification of the boundary stress tensor that we study in Section~\ref{sec:Junction}.

 \subsection{Equations of motion: Birkhoff theorem and black holes}
 With the explicit form of the two-dimensional action \req{eq:2daction} at hand, we can now compute the equations of motion of the Birkhoff-QT theories.  The full equations of motion in $D$ dimensions can be written as
 \begin{equation}
 \mathcal{E}_{ab}=8\pi G_{\rm N} T_{ab}\, ,
 \end{equation}
 where 
 \begin{equation}\label{Eabdef}
 \mathcal{E}_{ab}=\frac{16\pi G_{\rm N}}{\sqrt{|g|}}\frac{\delta S}{\delta g^{ab}}\, ,
 \end{equation}
 is the generalized Einstein tensor, and $T_{ab}$ is the stress energy tensor associated to matter
 \begin{equation}
 T_{ab}=-\frac{2}{\sqrt{|g|}}\frac{\delta S_{\rm matter}}{\delta g^{ab}}\, .
 \end{equation}
 The components of the equations of motion $\mathcal{E}_{ab}$ for spherically symmetric metrics \req{sphericmetric} can be obtained from the variation of \req{eq:2daction} with respect to $\gamma_{\mu\nu}$ and $\varphi$. By applying the chain rule of the functional derivative, we have
\begin{align}
\label{eq:eomtr}
\mathcal{E}_{\mu\nu}&=\frac{16\pi G_{\rm N}}{\Omega_{D-2} \varphi^{D-2}\sqrt{|\gamma|}}\frac{\delta S_{\rm 2d}}{\delta
\gamma^{\mu\nu}}\, ,\\
\label{eq:eomij}
\mathcal{E}_{ij}&=-\frac{g_{ij} 16\pi G_{\rm N}}{2(D-2)\Omega_{D-2} \varphi^{D-3}\sqrt{|\gamma|}}\frac{\delta S_{\rm 2d}}{\delta
\varphi}\, ,
\end{align}
where $i,j$ are the angular components and $\mu, \nu$ the two-dimensional components.  The explicit result from the variation of \req{eq:2daction} reads 
\begin{align}
 \mathcal{E}_{\mu\nu}&=\frac{D-2}{\varphi^{D-2}}\left[G_3 g_{\mu [\nu} \nabla_{\beta]} \partial^\beta \varphi  -\frac{G_2}{2} g_{\mu \nu} \right] \,,\\ \nonumber
 \mathcal{E}_{ij}&= \frac{1}{2 \varphi^{D-3}}\left[ \Box \varphi\, G_{3,\varphi}-G_{2,\varphi}-\frac{G_3}{2} R \right. \\& \left. -2 \nabla_{[a} \partial^b \varphi \nabla_{b]}  \partial^a   \varphi \, G_{3,X} \right] g_{ij}\,.
\end{align}
Observe that the angular components (equivalently, the equation for the scalar $\varphi$) are pure gauge, as they are related to the $\mu\nu$ components through the following Bianchi identity\footnote{We recall that $\nabla_{\mu}$ denotes the covariant derivative with respect to $\gamma_{\mu\nu}$.}
\begin{equation}
\nabla^\mu \left ( \varphi^{D-2} \mathcal{E}_{\mu \nu} \right) = \varphi^{D-3} \partial_\nu \, \varphi g^{ij} \mathcal{E}_{ij} \,.
\label{eq:bianchirel}
\end{equation}

Now, in order to make further progress, we can fix a particular gauge for the two dimensional metric $\gamma_{\mu\nu}$. Without loss of generality, we can put it into the form 
\begin{equation}
\mathrm{d}s_{\gamma}^2=-N(t,r)^2 f(t,r) \mathrm{d}t^2+\frac{\mathrm{d}r^2}{f(t,r)} \, ,
\label{eq:ssans}
\end{equation}
while we can also fix $\varphi=r$, which sets $X=f(t,r)$. In this case, the two-dimensional (time and radial) components of the generalized Einstein tensor read
\begin{align}
\mathcal{E}_{tt}&=\frac{(D-2) N^2 f}{2r^{D-2}} \frac{\partial}{\partial r} \left[r^{D-1}h(\psi) \right] \,,\\
\mathcal{E}_{tr}&=\frac{(D-2) r\partial_t \psi}{2f}  h'(\psi) \,  \,,\\
\mathcal{E}_{rr}&=\frac{(D-2) \partial_r N }{r N} h'(\psi) -\frac{1}{N^2 f^2}\mathcal{E}_{tt} \,,
\end{align}
where 
\begin{equation}
\psi=\frac{1-f(t,r)}{r^2}\, .
\end{equation}
On the other hand, the angular components $\mathcal{E}_{ij}$ can be derived from the Bianchi identity \eqref{eq:bianchirel}. As we anticipated, these equations are of second order and the differences with respect to GR are encapsulated in the function $h(\psi)$. 

In the presence of a stress-energy tensor $T_{ab}$, the dynamics of the system is driven by the equations
\begin{align}
 \frac{\partial }{\partial r} \left[r^{D-1}h(\psi) \right]&=\frac{16\pi G_{\rm N}}{(D-2)N^2 f} r^{D-2}T_{tt} \,,\\
\frac{\partial_t f}{r f}  &=-\frac{16\pi G_{\rm N}}{(D-2)h'(\psi)} T_{tr} \,  \,,\\
\frac{\partial_r N }{r N}&=\frac{8\pi G_{\rm N}}{(D-2)h'(\psi)} \left(T_{rr}+\frac{1}{N^2 f^2}T_{tt}\right) \,,
\end{align}
together with the conservation equation $\nabla^{a}T_{ab}=0$. These equations capture the full dynamical evolution of arbitrary spherically symmetric matter for infinite towers of Birkhoff gravities (or for finite subsets of those).

\subsubsection*{Vacuum solutions}
 
In vacuum, $T_{ab}=0$ and the equations above reduce to 
 \begin{equation}\label{vacuumeq}
 \partial_t f=0\, , \quad \partial_r N=0\, , \quad  \frac{\partial}{\partial r} \left[r^{D-1}h(\psi) \right]=0\, .
 \end{equation}
 Hence, $f=f(r)$ and $N=N(t)$, which can be reabsorbed in a redefinition of the time coordinate $N(t)^2 {\mathrm d}t^2 \rightarrow  {\mathrm d}t^2$. We thus conclude that the most general spherically symmetric solution of \req{QTaction} is in fact static and fully determined by a single function $f(r)$,
 \begin{equation}
\mathrm{d}s^2=- f(r) \mathrm{d}t^2+\frac{\mathrm{d}r^2}{f(r)}+r^2 \mathrm{d}\Omega^2_{D-2}\,.
\label{eq:ssans}
\end{equation}
The last equation in \req{vacuumeq} is integrated trivially and reduces to an algebraic equation for $f(r)$ 
 \be \label{eom}
h(\psi)  = \frac{2\mathsf{M}}{r^{D-1}}  \, ,
\ee
where  $\mathsf{M}$ is an integration constant which is related to the ADM mass of the solution, $M$, through 
\begin{equation}\label{newM}
\mathsf{M} \equiv \frac{8\pi G M}{(D-2)\Omega_{(D-2)}}\, . 
\end{equation}
This proves that our theories satisfy a Birkhoff theorem, extending previous results in the literature \cite{Oliva:2010eb, Oliva:2011xu, Cisterna:2017umf}. Next, we study in detail the properties of the solutions. 

\section{Regular black holes}
The solutions of \req{eom} are deformations of the Schwarzschild solution --- which is recovered when $\alpha_{n}=0$ $\forall n\ge 2$ --- and as shown by \cite{Bueno:2024dgm} they are singularity-free for very broad choices of the $\alpha_{n}$ couplings.\footnote{Depending on the values of the couplings, the solutions of \req{QTaction}  may also describe black holes with one or multiple horizons featuring curvature singularities at $r=0$ or finite-volume singularities --- see also \cite{Bueno:2024fzg,Bueno:2024qhh,Caceres:2024edr}.} The weakening of the singularity is generic even if the series is truncated at some $n=n_{\rm max}$, in which case one gets
\be 
f (r)= 1 - \left(\frac{2\mathsf{M} }{\alpha_{n_{\rm max} }} \right)^{1/n_{\rm max} } r^{2  - (D-1)/n_{\rm max} } + \cdots \, ,
\ee
near $r=0$.  When an infinite tower of corrections is included, the singularity is completely removed as long as the function $h(\psi)$ has an inverse for $\psi>0$ and the series that defines it, Eq.~\req{eom_psi}, has a finite radius of convergence. These conditions are met for very general choices of the $\alpha_n$ couplings , and for instance $\alpha_{n}(D-2n)\ge 0\,\, \forall\, n$ and $\lim_{n\rightarrow\infty} |\alpha_{n}|^{\frac{1}{n}}=C>0$ are sufficient conditions. 

We will consider some illustrative examples in order to discuss explicit results. The easiest possibility is to consider a set of couplings given by
\begin{equation}\label{exI}
\alpha^{(\rm I)}_n=\frac{(D-2)}{(D-2n)}\alpha^{n}\, ,\quad  h_{\rm I}(\psi)=\frac{\psi}{1-\alpha\psi}\, ,
\end{equation}
with $\alpha>0$, so that one gets the solution
\begin{equation}\label{Haywardf}
f_{\rm I}(r)=1-\frac{2\mathsf{M}r^2}{r^{D-1}+2\mathsf{M}\alpha}\, .
\end{equation}
This is nothing but the $D$-dimensional Hayward black hole, originally introduced in \cite{Hayward:2005gi} as a simple model of a regular black hole metric, but that in our case it appears as the unique spherically symmetric vacuum solution of a certain pure gravity theory with an infinite tower of higher-derivative corrections. 

Observe that this example only works in odd $D$. This is because for even $D$
the coupling with $n=D/2$ cannot be chosen as in \req{exI}. In fact, the term of order $n=D/2$ does not contribute to the series of $h(\psi)$ in \req{eom_psi}, and therefore one cannot achieve a summation of the form $h_{\rm I}$. This is just a technical nuance preventing us from finding certain simple summations of $h(\psi)$ in even dimensions, but it does not affect our conclusions. We further discuss this point in appendix \ref{ee}. 

Interestingly, the above theory \req{exI} can be extended to a broader family of theories given by a resummed characteristic polynomial 
\begin{align}\label{exN}
h_{\rm N}(\psi)=\frac{\psi}{\left( 1-\alpha^{\mathrm N}\psi^{\mathrm N}\right)^{1/\mathrm{N}}}\, ,
\end{align}
with ${\mathrm N}=1,2,3\ldots$ The values of the $\alpha_n^{(\mathrm N)}$s can be easily obtained in each case by Taylor-expanding $h_{\rm N}(\psi)$. The metric function reads
\begin{equation}\label{fN}
f_{\rm N}(r)=1-\frac{2\mathsf{M}r^2}{\left[ r^{\mathrm{N}(D-1)}+(2 \mathsf{M} \alpha)^{\mathrm{N}} \right]^{1/\mathrm{N}}}\, .
\end{equation}
For odd $D$, this provides an infinite family of theories and their corresponding regular black hole solutions for general values of ${\mathrm N} \in \mathbb{Z}^+$, including the Hayward case for $\mathrm{N}=1$. Whenever ${\mathrm N}$ is even, \req{exN} is always a valid theory also for $D=4(k+1)=8,12,16,\dots$, as in that case the problematic $n=D/2$ power is an even number, whereas only odd powers of $\psi$ appear in $h_{\mathrm N}(\psi)$. A simple interesting case belonging to this subclass was previously reported in \cite{Bueno:2024dgm} and it corresponds to $\mathrm{N}=2$, which yields a Bardeen-like metric \cite{1968qtr..conf...87B} (though it is not exactly the same).  On the other hand, whenever ${\mathrm N}$ is odd, both even and odd powers appear in the expansion of $h_{\mathrm N}(\psi)$ so not all cases exist for all $D=4(k+1)$. 
A sufficient condition for the existence of solutions in general even $D$ ---  namely, which also includes the cases $D=2(2k+1)=6,10,14,\dots$ --- is given by  ${\mathrm N}\geq D/2$. Hence, the simplest case which yields solutions valid in general even dimension is given by  ${\mathrm N}= D/2$,
\begin{equation}\label{fD2}
f_{D/2}(r)=1-\frac{2\mathsf{M}r^2}{\left[ r^{\frac{D(D-1)}{2}}+(2 \mathsf{M} \alpha)^{\frac{D}{2}} \right]^{2/D}}\, .
\end{equation}
 

An additional choice of couplings valid also for odd $D$ and $D=4(k+1)$ is given by
\begin{align}\notag 
 \alpha_n^{(\rm tanh)}&=\frac{(D-2)}{(D-2n)}\frac{((-1)^{(n+1)}+1)}{2n}\alpha^{n-1} \, ,\\ \label{tanhmodel}
   h_{\rm tanh}(\psi)=&\frac{{\rm arctanh}[\psi \alpha ]}{\alpha}\, .
 \end{align}
In this case, the solution is determined by
\begin{equation}
\label{tanhmodelfunc}
f_{\rm tanh}(r)=1- \frac{r^2}{\alpha} \tanh\left[\frac{2 \mathsf{M} \alpha}{r^{D-1}} \right]\, .
\end{equation}
For obvious reasons, we will refer to the theory defined by \eqref{tanhmodel} as the \emph{tanh model} and to \eqref{tanhmodelfunc} as the \emph{tanh black hole}. An interesting feature of this model is that the solution is $\mathcal{C}^{\infty}$ smooth but non-analytic at $r=0$, on account of terms of the form $e^{-3\mathsf{M}\alpha r^{-D+1}}$. 

\begin{figure*}
\centering
\includegraphics[width = 0.45\textwidth]{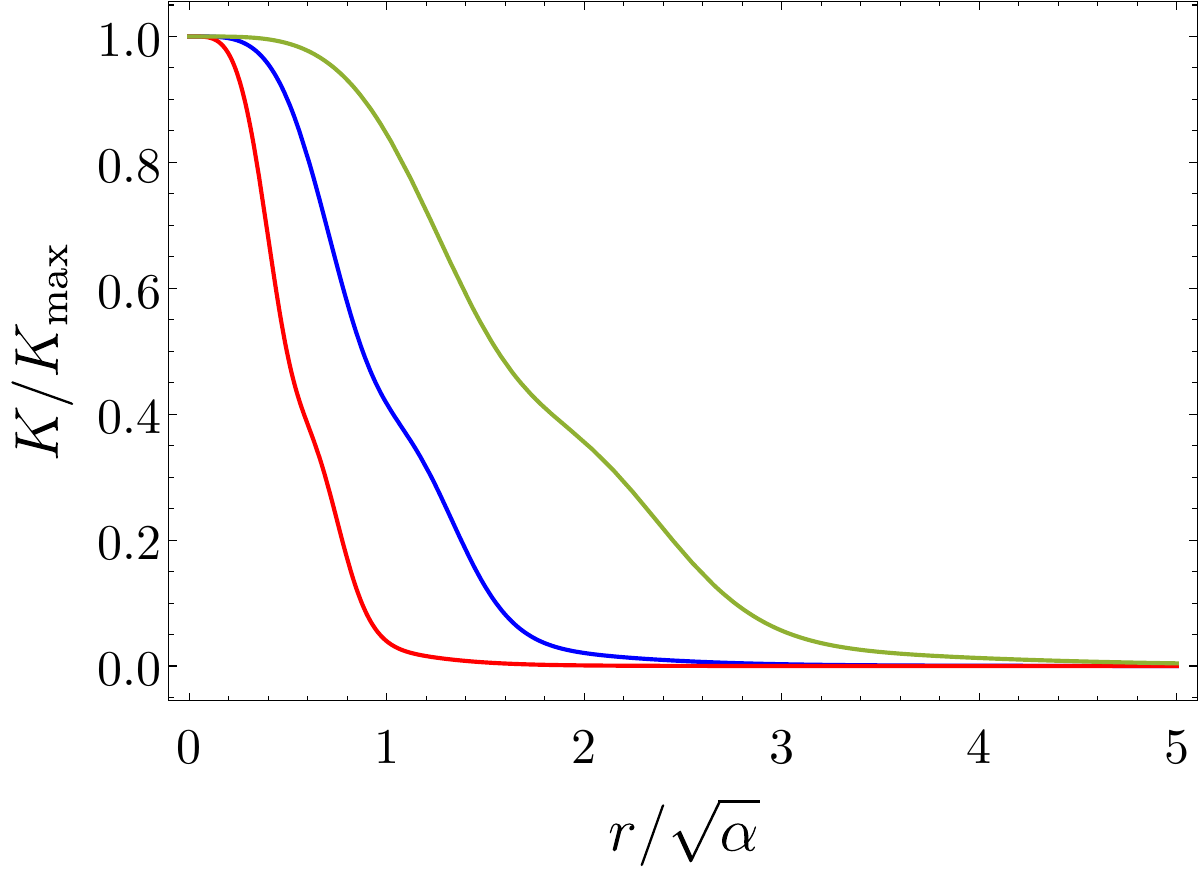}
\quad
\includegraphics[width = 0.45\textwidth]{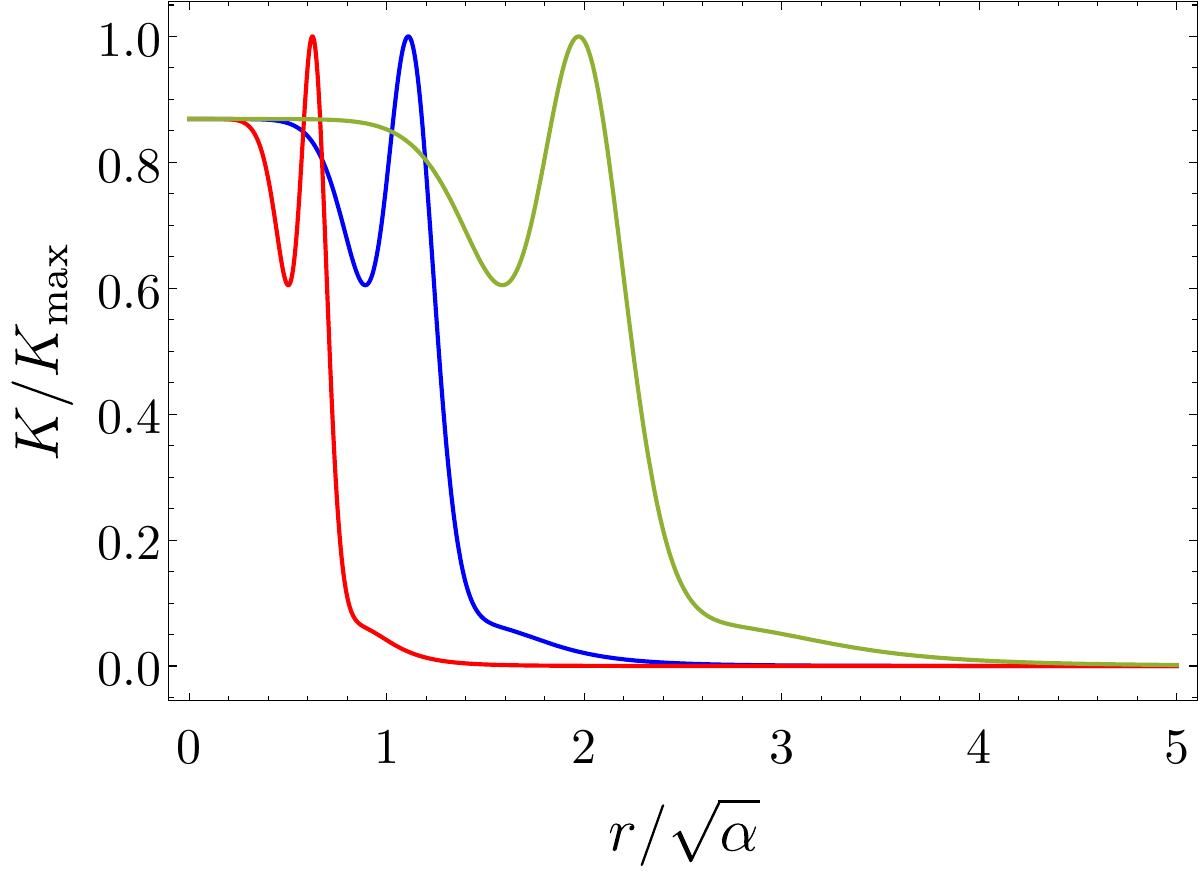}
\caption{{\it Left}: A plot of the Kretschmann scalar $K \equiv R_{abcd} R^{abcd}$ for the five-dimensional Hayward solution normalized its maximum value $K_{\rm max} = 40/\alpha^2$. The curves correspond to $\mathsf{M} = 0.1\mathsf{M}_{\rm cr}$ (red), $\mathsf{M} = \mathsf{M}_{\rm cr}$ (blue) and $\mathsf{M} = 10 \mathsf{M}_{\rm cr}$ (green). The plot illustrates a feature that holds for all masses and in all dimensions: There exists a maximum, solution-independent curvature that is achieved as $r \to 0$. {\it Right}: A plot of the Kretschmann scalar for the five-dimensional model \eqref{exN} with $\mathrm{N}=2$. The curves correspond to $\mathsf{M} = 0.1\mathsf{M}_{\rm cr}$ (red), $\mathsf{M} = \mathsf{M}_{\rm cr}$ (blue) and $\mathsf{M} = 10 \mathsf{M}_{\rm cr}$ (green). Here, we illustrate that there is again a universal, solution-independent bound on the curvature $K_{\rm max} \approx 46.0422/\alpha^2$. However, in this case the bound is reached at intermediate values of $r$.  
}
\label{fig:kretsch}
\end{figure*}

\subsection{Horizon structure}

Taking the Hayward metric as the prototypical example, let us examine the horizon structure of the solution. As usual, horizons are located at the zeros of $f(r)$. There is a critical value of the mass parameter which separates the different classes of solutions,
\be 
\mathsf{M}_{\rm cr}^{({\rm I})} \equiv \frac{(D-1)}{4} \left[\frac{\alpha (D-1)}{D-3} \right]^{\frac{(D-3)}{2}} \, .
\ee
When $\mathsf{M} > \mathsf{M}_{\rm cr}^{({\rm I})}$, there exists an event horizon $(r_{+})$ and an inner horizon ($r_{-}$). When $0 < \mathsf{M} < \mathsf{M}_{\rm cr}^{({\rm I})}$, there are no horizons. In this case, the solutions should be interpreted as gravitational solitons --- static, geodesically complete, finite energy Lorentzian metrics. 
When $\mathsf{M} = \mathsf{M}_{\rm cr}^{({\rm I})}$ the two horizons coincide and the metric has a degenerate Killing horizon. This critical spacetime is in some ways analogous to an extremal black hole --- it has vanishing Hawking temperature and an AdS$_2$ throat. However, in other ways, this critical spacetime is different from more familiar classical extremal black holes. First, these solutions do not mark the boundary between clothed and nakedly singular solutions. Second, there is no sense in which these critical solutions could be ``overcharged'' or ``overspun'', as there are no conserved charges at all, besides the mass, associated to them. Since the addition of positive energy can never drive a Hayward \textit{black hole} toward this extremal, or critical, limit, it seems the only mechanism to move toward the extremal limit would be via the quantum mechanical process of Hawking radiation.  Third, starting below the black hole threshold, {\it i.e.}~from one of the Hayward solitons, it seems that there would be nothing besides the pressure of matter to prevent one from forming an ``extremal'' horizon simply from the pile up of matter (as opposed to a charged or rotating case, where Coulomb repulsion or centrifugal forces would play an important role in hindering construction of the initial matter configuration). However, as we will return to in the discussion, there are reasons to believe that quantum effects will become important near the critical mass. It would be interesting to understand how the third law of black hole mechanics applies to these critical solutions, as it is not clear that it will do so in the same way as in General Relativity.

The solution space of the Hayward metric, {\it i.e.}, two-horizon black holes, critical/extremal solutions, and solitons is fully characteristic of all the models considered in this work. The only difference between models is the relationship between the critical mass parameter and the coupling constant, which differs depending on the chosen resummation --- for example, for the case given by \eqref{exN} with general ${\rm N}$ the critical mass parameter reads
\be 
M_{\rm cr}^{(\rm N)} = \frac{(D-1)^\frac{1}{{\rm N}}}{2^\frac{({\rm N} + 1)}{{\rm N}}} \left[ \frac{\alpha^{\rm N} (D-1)}{D-3}\right]^\frac{(D-3)}{2 {\rm N}}
\, . \ee
However, inclusion of matter or fine tuning of parameters can lead to even further possibilities, such as additional inner horizons or, interestingly, \textit{inner-extremal} regular black holes~\cite{DiFilippo:2024mwm}. 

\subsection{Interplay with the Limiting Curvature Hypothesis}

Long ago, Markov argued that a universal upper limit on the density of matter or the curvature of spacetime ought to be a fundamental physical principle~\cite{PismaZhETF.36.214}. There have since been many attempts to implement this limiting curvature hypothesis, and it has also been discussed in the context of regular black holes, {\it e.g.}~\cite{Frolov:2016pav, Frolov:2022fsl}. For example, it can be easily checked that the Hayward metric in any dimension has its curvature bounded from above by 
\be 
R_{abcd}R^{abcd} \le \frac{2 D (D-1)}{\alpha^2} \, ,
\ee
which is saturated as $r \to 0$. This bound is independent of the mass of the solution, and is completely fixed once the theory is specified via a value of $\alpha$. On the other hand, not \textit{all} regular black holes have such a fixed upper bound. In some cases, for example with the Bardeen metric,\footnote{Recall that the Bardeen metric function is~\cite{1968qtr..conf...87B} 
\begin{equation*}
f(r) = 1 - \frac{2 \mathsf{M} r^2}{\left(r^2 + \ell_0^2 \right)^{3/2}}.
\end{equation*} This geometry is not among those we have obtained as solutions of resummed quasi-topological theories.} the maximal curvature is related to the mass of the solution and hence can, in principle, grow without bound.

Does the limiting curvature hypothesis apply in our construction? Clearly it applies for certain resummations, \textit{e.g.}~the one resulting in the Hayward black hole. Furthermore, what can easily be proven is that the curvature at the core of the regular black holes is always solution-independent and universal. To see this, recall that for couplings $\alpha_{n}(D-2n)\ge 0$ satisfying $\lim_{n \to \infty} |\alpha_n|^{\frac{1}{n}} = C > 0$, $\psi_0 = C$ will correspond to the radius of convergence of the series defining $h(\psi)$. Due to the positivity of the coefficients, this means that $h(\psi)$ will diverge at $\psi = \psi_0$. Therefore, as $r \to 0$ we will have $\psi \to \psi_0$, indicating that the metric function behaves as $f \sim 1 - \psi_0 r^2$. Thus, the curvature at the core of the regular black hole will always have the limiting value
\be \label{R2psi0}
R_{abcd} R^{abcd} = 2 D (D-1) \psi_0^2 \, .
\ee
Since $\psi_0$ is simply the point where $h(\psi)$ diverges, it is a quantity that is specified purely by the theory and is universal in this sense --- see section \ref{gcons}.

Of course, the above argument does not prove that the curvature satisfies a universal upper bound everywhere in spacetime, as there could be intermediate regions of strong (but finite) curvature. Assessing this is a much more involved problem. We have studied it on a case-by-case basis for the resummations presented in~\cite{Bueno:2024dgm}, along with the \textit{tanh black hole} constructed in this work. In each of these cases, we observe that there is a universal, solution-independent upper bound on the curvature, though it does not always coincide with the value of the curvature at $r = 0$. One such example where this occurs is the solution \eqref{fN} introduced above. We illustrate this example with $\mathrm{N}=2$, along with the Hayward one for comparison, in Figure~\ref{fig:kretsch}, both in five dimensions. Here, we show the Kretschmann scalar as a function of radius for regular solutions of different masses (indicated by the different coloured curves). In the case of the solution~\eqref{fN} with $\mathrm{N}=2$ (the so-called Bardeen-like solution), we see that there is a solution-independent, universal upper bound on the curvature $R_{abcd}R^{abcd} \approx 46.0422/\alpha^2$. However, in contrast to the Hayward black hole, this universal upper bound is achieved away from $r = 0$. 

Proving the existence of an upper bound in general, or establishing necessary and sufficient conditions on the resummations, would be a challenging but interesting problem --- see~\cite{Frolov:2024hhe} for related considerations.

\section{Boundary terms and junction conditions}\label{sec:Junction}

Assume we consider our theory \eqref{QTaction} on a $D$-dimensional manifold $\mathcal{M}$ with a certain domain wall $\Sigma$ that splits the spacetime into two manifolds $\mathcal{M}_+$ and $\mathcal{M}_-$. The hypersurface $\Sigma$ has a certain surface stress-energy tensor $S_{AB}$, and we wish to solve the modified Einstein equations in the presence of this discontinuous matter distribution. The best way to achieve this is through generalized Israel junction conditions, that we derive next. 
 
 The first junction condition takes the same form as in GR, and it simply states that the induced metric, $h_{AB}$, is continuous across the hypersurface, 
 \begin{equation}
 h_{AB}^{+}=h_{AB}^{-}\, ,
 \end{equation}
 where $h_{AB}^{+}$ and $h_{AB}^{-}$ represent the induced metric computed at each side of the boundary (we use capital Latin indices to refer to boundary indices). 
 
The second junction condition does depend on the modified Einstein's equations, and the best way to write it down is through the analysis of the boundary terms in the variation of the action. To this end, we would need to supplement the action \req{QTaction} with a generalized York-Gibbons-Hawking boundary term,
\begin{equation}\label{Stotal1}
S^{\rm total}=S+S^{\rm boundary}\, ,
\end{equation}
that makes the variational problem well posed. The total variation of this action then would read
\begin{equation}\label{variationStotal}
\begin{aligned} 
\delta S^{\rm total}&=\frac{1}{16\pi G_{\rm N}}\int \mathrm{d}^{D}x\sqrt{|g|}\mathcal{E}_{ab}\delta g^{ab}\\
&+\frac{1}{16\pi G_{\rm N}}\int_{\Sigma} \mathrm{d}^{D-1}x\sqrt{|h|}\Pi_{AB}\delta h^{AB}\, ,
\end{aligned}
\end{equation}
where $\mathcal{E}_{ab}$ are the equations of motion as defined in \req{Eabdef}, and $\Pi_{AB}$ represents the boundary equations of motion, which we can also write as
\begin{equation}
\Pi_{AB}=\frac{16\pi G_{\rm N}}{\sqrt{|h|}}\frac{\delta S^{\rm total}}{\delta h^{AB}}\, .
\end{equation}
The second junction condition is written in terms of this tensor; it states that the discontinuity of $\Pi_{AB}$ is given by the surface stress-energy tensor~\cite{Israel:1966rt, Brown:1992br, Davis:2002gn, Padilla:2012ze},
\begin{equation}\label{2ndjunction}
\Pi^{-}_{AB}-\Pi^{+}_{AB}=8\pi G_{\rm N} S_{AB}\, .
\end{equation}
Now, the main difficulty lies in finding the boundary terms in \req{Stotal1}, since the variational problem in higher-derivative theories has many issues. However, if the spacetime is spherically symmetric, we can use the two-dimensional action \req{eq:2daction}, which allows for a well-posed variational problem since it is a Horndeski theory and has second-order equations of motion. 

Take a unit normal vector $n_a$ to $\Sigma$ which is normalized as $n_{a}n^{a}=\epsilon=\pm 1$. 
If $\Sigma$ is spherically symmetric, we can parametrize it by 
 \begin{equation}
 h_{AB}\mathrm{d}x^{A}\mathrm{d}x^{B}=h_{\tau\tau}\mathrm{d}\tau^2+\varphi(\tau)^2 \mathrm{d}\Omega_{D-2}^2\, .
 \end{equation}
From the point of view of the two-dimensional metric $\gamma_{\mu \nu}$ (see equation \req{sphericmetric}), $\Sigma$ is just a curve with induced metric $ds_{h}^2=h_{\tau\tau} d\tau^2$.  We can always go to a gauge in which $h_{\tau\tau}=-\epsilon=\mp1$, so that $\tau$ represents the proper time/length of the curve, but we will keep $h_{\tau\tau}$ general for now, since we need to analyze the variational problem. 

The action \eqref{eq:2daction} contains second-derivative terms and its variation will produce terms including the variation of the normal derivatives of $\varphi$ and of the induced metric on the curve delimiting the domain wall. As a consequence, it is necessary to include boundary terms to fix this aspect. 
The boundary terms for general Horndeski gravity were analyzed by \cite{Padilla:2012ze}, so we refer to that work for details about the computation. We have independently computed the boundary terms for the case of the two-dimensional Horndesky theory \req{horn}, finding perfect agreement. The total action with boundary terms reads
\begin{align}
\label{eq:actot2d}
S_{\rm 2d}^{\rm total}&=S_{\rm 2d}+ \frac{(D-2) \Omega_{D-2}}{16 \pi G_{\rm N}}B\,,\\
B&=\int_\Sigma \mathrm{d}\tau \sqrt{\vert h_{\tau\tau} \vert} \left[ F_3+2G_4 K +4 \,\Box^h \varphi\,  F_{4,Y}\right] \,,
\end{align}
where $\varphi_n=n^\mu \partial_\mu \varphi$, $K=\nabla_\mu n^\mu$ is the extrinsic curvature of the curve $\Sigma$, $\Box^h \varphi$ is the Laplacian on $\Sigma$ and
\begin{equation}
 F_{l}=\int_{0}^{\varphi_n}  G_{l}\left(\varphi, Y+\epsilon z^2\right)\, \mathrm{d}z\, , \quad Y=h^{\tau\tau} \dot\varphi^2\,,
\end{equation}
where $l=3,4$, $\dot \varphi=\mathrm{d}\varphi/\mathrm{d}\tau$, and $h^{\tau\tau}=1/h_{\tau\tau}$. 

The variation of this action takes the schematic form
\begin{equation}
\begin{aligned}
\delta S^{\rm total}_{\rm 2d}&=\int \mathrm{d}^{2}x\left[\frac{\delta S_{\rm 2d}}{\delta
\gamma^{\mu\nu}}\delta \gamma^{\mu\nu}+\frac{\delta S_{\rm 2d}}{\delta
\varphi}\delta \varphi\right]\\
&+\int_{\Sigma} \mathrm{d}\tau \sqrt{|h_{\tau\tau}|}\left[J_{\tau\tau}\delta h^{\tau\tau}+J_{\varphi}\delta\varphi\right]\, ,
\end{aligned}
\end{equation}
and comparison with \req{variationStotal} leads, on the one hand, to the relationships \eqref{eq:eomtr}, \eqref{eq:eomij}, and on the other, to the identification of the components of the $(D-1)$-dimensional boundary tensor $\Pi_{AB}$, 
\begin{align}
\label{eq:eombtr}
\Pi_{\tau \tau}&=\frac{16\pi G_{\rm N}}{\Omega_{D-2} \varphi^{D-2}}J_{\tau\tau}\, ,\\
\label{eq:eombij}
\Pi_{ij}&=-\frac{g_{ij} 16\pi G_{\rm N}}{2(D-2)\Omega_{D-2} \varphi^{D-3}}J_{\varphi}\, .
\end{align}
The variation of the two-dimensional action \eqref{eq:actot2d} and the extraction of $J_{\tau\tau}$ and $J_{\varphi}$ is quite tedious, so we present here the final result. We find
\begin{align}
\label{eq:eom2btr}
\Pi_{\tau\tau}&=\frac{(D-2)\epsilon}{2\varphi^{D-2}}F_{3}\,, \\
\label{eq:eom2bij}
\Pi_\varphi \equiv g^{ij} \Pi_{ij} &=-\frac{\epsilon}{\varphi^{D-3}\dot \varphi} \frac{\mathrm{d}}{\mathrm{d} \tau} \left ( \varphi^{D-2} \Pi_{\tau \tau} \right)\,,
\end{align}
where we have imposed $h_{\tau\tau}=-\epsilon$ so that $\tau$ is the proper time/length. 
These expressions match exactly those of \cite{Padilla:2012ze} when restricting to our particular choices of functions $G_2, G_3$ and $G_4$ in the 2-dimensional theory given by \eqref{eq:actot2d}. We observe that all the relevant information is encoded in $\Pi_{\tau \tau}$. Using \req{G3form}, this gets simplified into
 \begin{equation}\label{Pitautau}
 \Pi_{\tau\tau}=\frac{(D-2)\epsilon}{\varphi}\int_{0}^{\varphi_n} \mathrm{d}z h'\left(\frac{1+\epsilon \dot\varphi^2-\epsilon z^2}{\varphi^2}\right)\, .
 \end{equation}

\section{Thin-shell collapse}

\subsection{The shell equation}

Let us consider a thin spherical shell of ``dust'' (that is, presureless matter), with a surface stress-energy tensor given by
\begin{equation}
S_{AB}=\sigma u_{A} u_{B}\, ,
\end{equation}
where $\sigma$ is the surface matter density and $u_{A}$ is the $(D-1)$-velocity field of the fluid defined on $\Sigma$. If we parametrize the shell's hypersurface by its proper time $\tau$, then the components of $S_{AB}$ are simply
\begin{equation}
S_{\tau\tau}=\sigma\, ,\quad S_{ij}=0\, .
\end{equation}
The radius of the shell is denoted by $R(\tau)$. On account of Birkhoff's theorem, the interior region $r< R(\tau)$ is flat spacetime, while the exterior region $r>R(\tau)$ must correspond to the unique metric \req{eq:ssans} for some value of the mass $M$, 
\begin{align}
\mathrm{d}s_{-}^2&=-\mathrm{d}t_{-}^2+\mathrm{d}r^2+r^2\mathrm{d}\Omega_{D-2}^2\, ,\\
\mathrm{d}s_{+}^2&=- f(r) \mathrm{d}t_{+}^2+\frac{\mathrm{d}r^2}{f(r)}+r^2 \mathrm{d}\Omega^2_{D-2}\,,
\end{align}
where the time coordinates $t_{\pm}$ are in principle different in each region. Our goal is to determine the evolution of the shell radius $R(\tau)$.  To this end, we make use of the junction conditions. We parametrize the shell's hypersurface in each side by
\begin{equation}
(t_{\pm},r)=(T_{\pm}(\tau),R(\tau))\, ,
\end{equation}
so that the induced metric reads
\begin{align}
\mathrm{d}s_{\Sigma-}^2&=-\mathrm{d}\tau^2\left(\dot{T}_{-}^2-\dot{R}^2\right)+R(\tau)^2 \mathrm{d}\Omega^2_{D-2}\, ,\\
\mathrm{d}s_{\Sigma+}^2&=-\mathrm{d}\tau^2\left(f(R)\dot{T}_{+}^2-\frac{\dot{R}^2}{f(R)}\right)+R(\tau)^2 \mathrm{d}\Omega^2_{D-2}\, .
\end{align}
The first junction condition requires the continuity of the induced metric, and furthermore, since $\tau$ is the proper time, it must be equal to
\begin{align}
\mathrm{d}s_{\Sigma-}^2=\mathrm{d}s_{\Sigma+}^2=-\mathrm{d}\tau^2+R(\tau)^2 \mathrm{d}\Omega^2_{D-2}\, .
\end{align}
Therefore, we get
\begin{align}\label{betaminus}
\dot{T}_{-}&=\sqrt{1+\dot{R}^2}\equiv \beta_{-}\, ,\\
f(R)\dot{T}_{+}&=\pm\sqrt{f(R)+\dot{R}^2}\equiv \beta_{+}\, ,
\label{betaplus}
\end{align}
following the terminology of Poisson \cite{Poisson:2009pwt}. Let us note that, while $\beta_{-}$ is always taken to be positive, the sign of $\beta_{+}$ can change in the black hole interior. The rule is that $\beta_{+}$ starts being positive in the exterior region, and the sign  in front of the square root in \req{betaplus} must be flipped whenever we reach a point with $\beta_{+}=0$.   

Then we have to impose the second junction condition \req{2ndjunction}. For the evaluation of $\Pi_{\tau\tau}$ in \req{Pitautau}, we take into account that the normal vector is spacelike $(\epsilon=+1)$ and given by
\begin{align}
(n^{-})^{\mu}\partial_{\mu}&=\dot{R}\partial_{t_{-}}+\dot{T}_{-} \partial_{r}\, ,\\
(n^{+})^{\mu}\partial_{\mu}&=\frac{1}{f(R)}\dot{R}\partial_{t_{+}}+f(R)\dot{T}_{+} \partial_{r}\, ,
\end{align}
on each side of the shell. Then, using that $\varphi=r$, we get
\begin{equation}
\varphi_{n}^{\pm}\equiv (n^{\pm})^{\mu}\partial_{\mu}\varphi=\beta_{\pm}\, ,
\end{equation}
where we also used \req{betaplus}, \req{betaminus}. 
Therefore, 
 \begin{equation}\label{Pitautau}
 \Pi_{\tau\tau}^{\pm}=\frac{(D-2)}{R}\int_{0}^{\beta_{\pm}} \mathrm{d}z h'\left(\frac{1+ \dot{R}^2-z^2}{R^2}\right)\, ,
 \end{equation}
 and the junction conditions read
 \begin{align}\label{junctiontau}
\Pi_{\tau\tau}^{-}-\Pi_{\tau\tau}^{+}&=8\pi G_{\rm N}\sigma\, ,\\
\frac{d}{d\tau}\left[R^{D-2}\left(\Pi_{\tau\tau}^{-}-\Pi_{\tau\tau}^{+}\right)\right]&=0\, ,
\label{junctionphi}
 \end{align}
where in the second one, coming from the angular components $\Pi_{ij}^{-}-\Pi_{ij}^{+}=0$, we used \req{eq:eom2bij}. From the combination of \req{junctiontau} and \req{junctionphi}, we immediately conclude that the proper mass of the shell, $m$, is constant
\begin{equation}
m\equiv \sigma R^{D-2}\Omega_{D-2}=\text{constant}. 
\end{equation}
Using this result, we can write \req{junctiontau} as
\begin{equation}\label{shellequation1}
\frac{\mathsf{m}}{R^{D-3}}=\int_{\beta_{+}}^{\beta_{-}} \mathrm{d}z h'\left(\frac{1+ \dot{R}^2-z^2}{R^2}\right)\, ,
\end{equation}
where, in analogy with \req{newM}, we have introduced the mass parameter
\begin{equation}\label{newlittlem}
\mathsf{m}=\frac{8\pi G_{\rm N} m}{(D-2)\Omega_{D-2}}\, .
\end{equation}

\subsubsection*{A theory-independent form of the shell equation}

As we can see, the theory dependence is encoded in the $h'(x)$ function that we need to integrate, and also implicilty in $f(R)$ that enters in $\beta_{+}$. We can massage \req{shellequation1} to make the explicit dependence on $h(x)$ disappear. To achieve this, we perform a change of variables $z=z(r)$ defined by
\begin{equation}\label{eq:zeq}
h\left(\frac{1+ \dot{R}^2-z^2}{R^2}\right)=\frac{2\mathsf{M}}{r^{D-1}}\, .
\end{equation}
Differentiating both sides and dividing by $-2z/R^2$, we have
\begin{equation}
\mathrm{d}z h'\left(\frac{1+ \dot{R}^2-z^2}{R^2}\right)=\frac{R^2(D-1)\mathsf{M}}{z r^{D}}\mathrm{d}r\, .
\end{equation}
On the other hand, since \req{eq:zeq} is nothing but the equation of $f(r)$ \req{eom}, we conclude that $z$ is given by
\begin{equation}
z=\sqrt{1+\dot{R}^2-\frac{R^2}{r^2}\left[1-f(r)\right]}\, .
\end{equation}
Finally, in the case $\beta_{+}>0$, the limits of integration are mapped to $z=\beta_{+}\Rightarrow r=R$, $z=\beta_{-}\Rightarrow r\to \infty$. Therefore, we rewrite  the shell equation \req{shellequation1} as
\begin{equation}\label{sdd}
\frac{\mathsf{m}}{R^{D-1}}=\int_{R}^{\infty} \frac{\mathrm{d}r \, \mathsf{M} (D-1)}{r^D \sqrt{1+\dot R^2-\frac{R^2}{r^2}\left[1-f(r) \right]}}\, .
\end{equation} 
This equation is valid for any QT theory of the class considered in \req{QTaction}, and remarkably it has a theory-independent form, although of course $f(r)$ must be the solution of the corresponding theory. 

In the case $\beta_{+}\le 0$, the result is slightly more complicated. We get the same integrand, but the domain of integration contains two intervals, corresponding to $z>0$ and $z<0$. The point $z=0$ happens at a radius $r_{0}$ given by
\begin{equation}
r_{0}=\left[\frac{2\mathsf{M}}{h\left(\frac{1+\dot{R}^2}{R^2}\right)}\right]^{1/(D-1)}\, ,
\end{equation}
and one can see that the equation \req{shellequation1} becomes in this case
\begin{equation}\label{sdd2}
\frac{\mathsf{m}}{R^{D-1}}=\int_{R}^{\infty} \frac{\mathrm{d}r \, \mathsf{M} (D-1)\operatorname{sign}\left(r-r_{0}\right)}{r^D \sqrt{1+\dot R^2-\frac{R^2}{r^2}\left[1-f(r) \right]}}\, .
\end{equation} 
Plugging the black hole metric function $f(r)$ for the chosen model either in \req{sdd} or \req{sdd2}  yields an integro-differential equation for $R(\tau)$. In general, we find that only one of the equations admits for a real solution, and it corresponds to \req{sdd} if $\beta_{+}>0$ and  \req{sdd2} if $\beta_{+}<0$, where the sign of $\beta_{+}$ must be flipped in the way we explained below \req{betaplus}.  The transition from one equation to another is however completely smooth, and it gives a smooth differential equation for $R(\tau)$.  

\subsubsection*{The shell equation as an energy integral}
If instead of \req{eq:zeq}, we do the change of variable
\begin{equation}
h\left(\frac{1+ \dot{R}^2-z^2}{R^2}\right)=\frac{2E}{R^{D-1}}\, ,
\end{equation}
and we use $E$ as a parameter, the integral \req{shellequation1} then takes the form
\begin{equation}\label{sdd3}
\mathsf{m}=\int_{0}^{\mathsf{M}}\frac{\mathrm{d}E}{\sqrt{\dot{R}^2+f(R,E)}}\, ,
\end{equation}
where $f(R,E)$ is the solution with mass parameter $\mathsf{M}=E$, and we are assuming $\beta_{+}\ge 0$. Again, this formula looks theory-independent, but the theory dependence is encoded in the form of $f(R,E)$.

\subsection{Solution of the shell equation}
The shell equation, expressed in either of its forms, \req{shellequation1}, \req{sdd} or \req{sdd3}, defines implicitly a differential equation for $R(\tau)$, of the form $F(R,\dot{R})=0$.  It is convenient to recast this differential equation in the form
\begin{equation}\label{tse}
\dot R^2+V(R)= \frac{\mathsf{M}^2}{\mathsf{m}^2}-1 \, , 
\end{equation}
where $V(R)$ is an effective potential. The equations \req{shellequation1}, \req{sdd} or \req{sdd3} then become an equation for the potential $V(R)$.  We have pulled out the factor $\frac{\mathsf{M}^2}{\mathsf{m}^2}-1$, which is equivalent to an energy level, advancing that $V(R)\to 0$ for $R\to\infty$.

\begin{figure*}
\centering \hspace{0cm}
\includegraphics[width=0.49\textwidth]{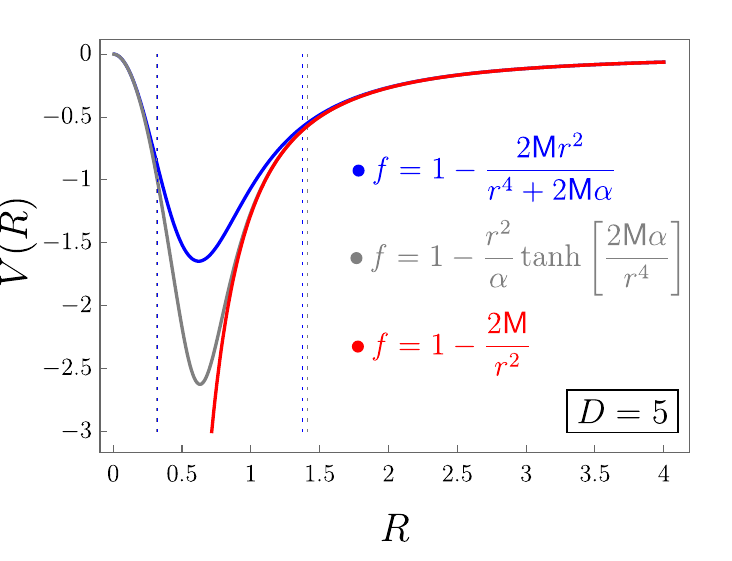}
\includegraphics[width=0.49\textwidth]{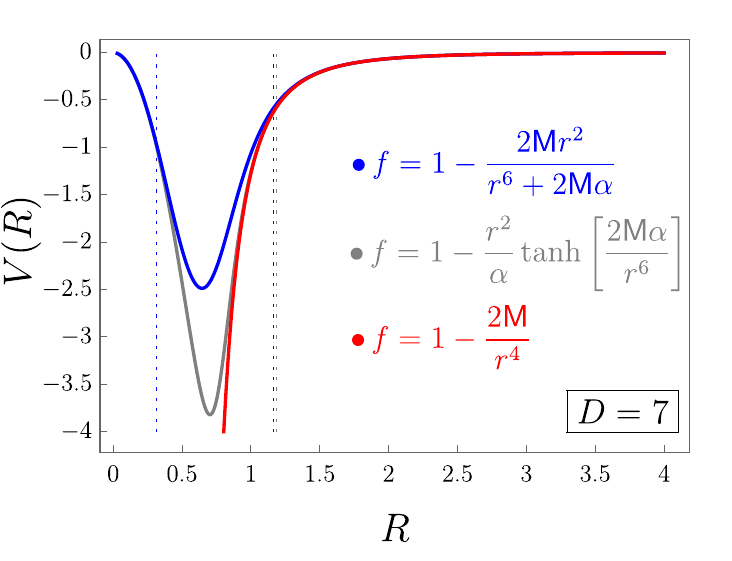}
\caption{We plot the thin shell effective potential as defined in \req{tse} in $D=5$ (\emph{Left}) and $D=7$ (\emph{Right}) respectively for two Quasi-topological theories whose vacuum SS solutions correspond to a Hayward black hole with metric function \req{Haywardf} (blue) and a {\emph{tanh}} black hole with metric function \req{tanhmodelfunc} (gray). The dashed lines correspond to the respective locations of the inner and outer horizons. The Einstein gravity shell potential, given by \req{vRE}, for which the exterior solution is a Schwarzschild black hole is shown in red.  In all cases we set $ {\mathsf M}=1$, ${\mathsf m}=1.05$,  $\alpha=1/10$.  }
\label{fig:VR}
\end{figure*}

\subsubsection{Einstein gravity}
In the particular case of Einstein gravity, in which $h(\psi)=\psi$, the integral \req{shellequation1} is straightforward and we get 
\begin{equation}
\frac{\mathsf{m}}{R^{D-3}}=\beta_{-}-\beta_{+}\, .
\end{equation}
Solving this equation for $\dot{R}^2$ and further using that $f(r)=1-2\mathsf{M}/r^{D-3}$, we obtain precisely \req{tse} with 
\begin{equation}\label{vRE}
V(R)=-\frac{\mathsf{M}}{R^{(D-3)}}-\frac{\mathsf{m}^2}{4R^{2(D-3)}}\, ,
\end{equation}
which is a monotonously decreasing function of $R$ as one moves towards $R=0$. Starting at any finite radius $R(0)=R_0$, the shell collapses leaving behind a Schwarzschild black hole and reaching $R=0$ after a finite proper time. In particular, near $R=0$ the time evolution of the shell is $R(\tau)\sim (\tau_0-\tau)^{1/(D-2)}$, which cannot be extended for $\tau>\tau_{0}$, signaling the breakdown of GR. 

\subsubsection{Hayward black hole}
As a first illustrative case, we consider the Hayward black hole \req{Haywardf}, which we found to be an exact solution of odd-dimensional theories given by the choice of couplings \req{exI}. We note that in odd dimensions this metric is not only regular --- in the sense of having finite curvature everywhere --- but completely smooth ($\mathcal{C}^{\infty}$). This is important in order to avoid some issues associated to non-smoothness at $r=0$ \cite{Zhou:2022yio}.   

The integration in \req{shellequation1} can be carried out explicitly and we obtain the following equation 
\begin{widetext}
\begin{equation}\label{HaywardShellEq}
\mathsf{m}=\frac{R^{D-1}}{2 \left(R^2-\alpha\beta_{-}^2\right)}  \left(\beta_{-}-\left(1+\frac{2\alpha\mathsf{M}}{R^{D-1}}\right)\beta_{+}+\frac{R^2\arctan\left(\frac{\sqrt{\alpha} (\beta_{-}-\beta_{+})\sqrt{R^2-\alpha \beta_{-}^2}}{R^2-\alpha\beta_{-}(\beta_{-}-\beta_{+})}\right)}{\sqrt{\alpha} \sqrt{R^2-\alpha\beta_{-}^2}}\right)\, .
\end{equation}
\end{widetext}
This is too complicated to be solved analytically, but we can obtain information about it by solving it in different regimes.  Near infinity $R\to\infty$, the corrections to GR are small and we can expand the solution in powers of $\alpha$ (which is essentially equivalent to an expansion in $1/R^{D-3}$). We get in this case
\begin{equation}
\begin{aligned}
V(R)&=-\frac{\mathsf{M}}{R^{(D-3)}}-\frac{\mathsf{m}^2}{4R^{2(D-3)}}\\
&+\frac{\alpha}{R^2}\left(\frac{4\mathsf{M}^2}{3R^{2(D-3)}}+\frac{ \mathsf{M} \mathsf{m}^2}{R^{3(D-3)}}+\frac{\mathsf{m}^4}{6R^{4(D-3)}}\right)\\
&+\mathcal{O}(\alpha^2)\, ,
\end{aligned}
\end{equation}
and we observe that in fact the corrections do not affect the asymptotic value of $V(R)$.  

It is most interesting to understand the behavior of the solution for small $R$, where we expect the corrections to be important. Remarkably, it is also possible to obtain an analytic expression in such a regime for $V(R)$ (which is in practice equivalent to an expansion in $1/M$), and we get that the potential takes the form
\begin{equation}\label{eq:VHayward0}
V(R)=-\frac{R^2}{\alpha}-\frac{R^{D+1} \log \left (\frac{R^{D-1}}{2\alpha \mathsf{M}} \right) }{2\alpha^2 \mathsf{M}}+\mathcal{O}(R^{D+3})\,.
\end{equation}
The expansion of $V(R)$ features logarithmic terms, but is fully regular and vanishes at $R=0$.

The full shape of $V(R)$ can be obtained by solving \req{HaywardShellEq} numerically, and we show it in Fig.~\ref{fig:VR} in the $D=5$ and $D=7$ cases. 
From this plot we can easily understand the motion of a collapsing shell.

\begin{figure*}
\centering \hspace{0cm}
\includegraphics[width=0.49\textwidth]{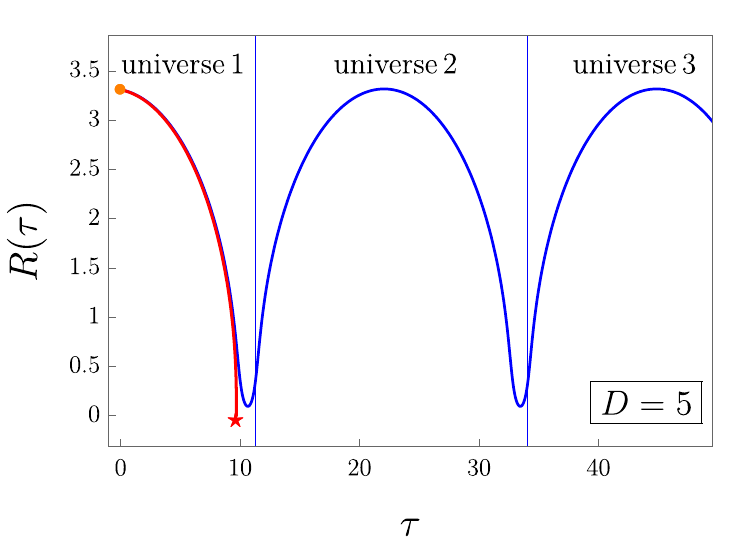}
\includegraphics[width=0.49\textwidth]{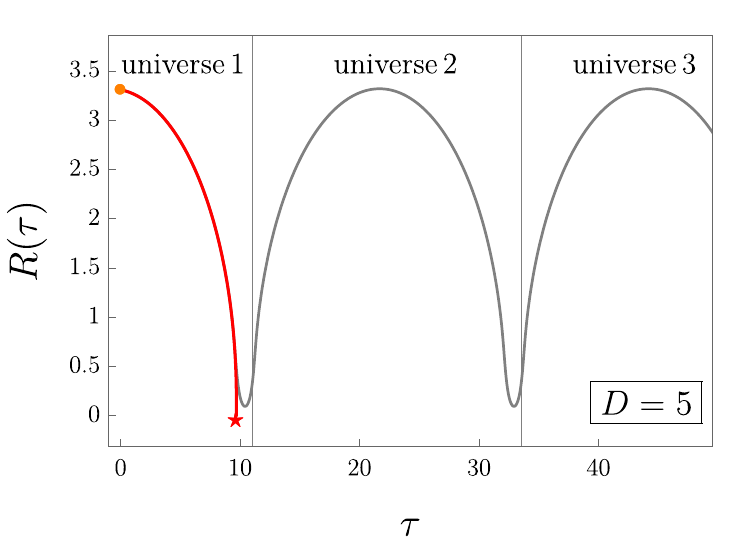}\\ 
\includegraphics[width=0.49\textwidth]{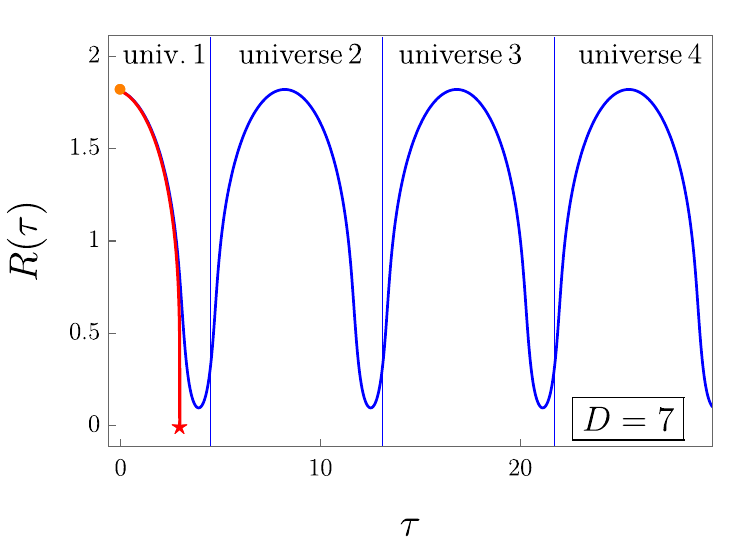}
\includegraphics[width=0.49\textwidth]{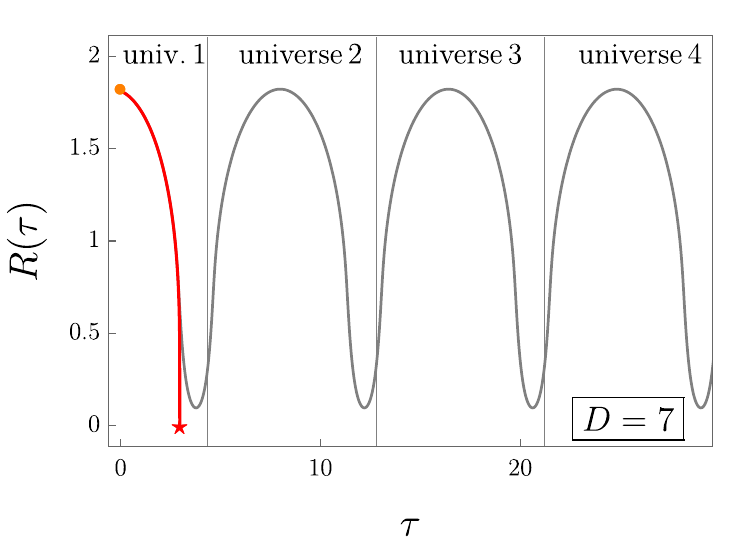}
\caption{We plot the shell radius as a function of the proper time for five-dimensional (\emph{upper row}) and seven-dimensional (\emph{lower row}) theories. In each case, the blue curve corresponds to a Quasi-topological theory whose vacuum SS solution is a Hayward black hole with metric function \req{Haywardf} and the gray one to a theory whose vacuum SS solution is given by a {\emph{tanh}} black hole with metric function \req{tanhmodelfunc}. The Einstein gravity evolution --- which ends at a singularity (marked with a red star) after finite proper time --- is displayed in red in all plots. The vertical lines represent the points at which the shells cross the black hole inner horizons from the inside, appearing in a new universe.   In all cases we set $ {\mathsf M}=1$, ${\mathsf m}=1.05$,  $\alpha=1/10$.    }
\label{fig:Rtau}
\end{figure*}

(1) If the shell starts at rest at some radius $R_{0}$, then we have $M<m$, and the motion is bounded. If $R_0$ is larger than the radius at which the potential reaches its minimum, the shell will roll down the potential towards lower values of $R$, it will cross the minimum of the potential, and then it will reach a turning point $R_{\rm min}>0$ at which $\dot{R}=0$. Here, the shell bounces and starts to expand again, reaching the starting point $R_{0}$, where the trajectory restarts in a cyclic way --- see Fig.\,\ref{fig:Rtau}. 

(2) If the shell starts from rest rest at infinity, then $M=m$, and from the point of view of \req{tse}, the shell moves on a ``zero energy'' trajectory. Thus, the shell will move towards smaller radii, and it will climb the potential up to $R=0$, which is only reached in an infinite time. In fact, using \req{eq:VHayward0}, we can see that the motion near $R=0$ is given by $R(\tau)\sim e^{-\tau/\sqrt{\alpha}}$. Thus, the shell would tend to stay at $R=0$, but this is an unstable equilibrium and any small deviation will make the shell roll again from the $R=0$ towards infinity. 

(3) Finally, assume that $M >m$. In that case, the shell starts collapsing with a negative radial velocity at infinity and it will reach $R=0$ with finite kinetic energy, so that the shell will cross $R=0$ in a smooth way.  Using again \req{eq:VHayward0}, one can see that the motion of the shell is given by
\begin{equation}
R(\tau)=|\tau_0-\tau|\sqrt{\frac{\mathsf{M}^2}{\mathsf{m}^2}-1}\left(1+\frac{(\tau_0-\tau)^2}{6\alpha}+\ldots\right)
\end{equation}
Since the dust particles do not interact, they can naturally cross each other, and the motion is smooth in the sense that each particle follows a smooth trajectory, where after crossing $R=0$ each point in the shell has to be identified with its antipodal one.  Interestingly, the logarithmic terms in \req{eq:VHayward0} introduce a small non-smoothness at $R=0$, as the trajectory is actually $\mathcal{C}^{D}$ at that point. However, this is acceptable as this is higher than the order of the equations of motion.

We note that, depending on the total mass $\mathsf{M}$, the collapse will give rise to either a horizon, an extremal horizon, or no horizon at all,  but the motion of the shell is qualitatively the same in all cases --- obviously, the causal structure of the spacetime is not, as we analyze below. It may come as a bit of a surprise that gravitational collapse can produce an extremal horizon. However, we note that precisely the same thing happens with extremal charged shells in Einstein-Maxwell theory~\cite{IsraelBounce}. If $\mathsf{M}$ is large enough to form a horizon, it can be proven that the turning point always occurs inside the inner horizon  $R_{\rm min}<r_{-}$.

\subsubsection{Tanh black hole}

Another interesting example in which to examine the dynamical collapse of a spherical thin shell is provided by the \emph{tanh} model presented in \eqref{tanhmodel}. In this case, the expression \eqref{shellequation1} can be fully integrated into
\begin{widetext}
\begin{equation}
\mathsf{m}=\frac{R^{D-1}}{2\sqrt{\alpha}} \left [\frac{\arctan \left (\frac{\sqrt{\alpha R^2 -\alpha^2 \beta_-^2}(\beta_--\beta_+)}{R^2+\alpha\beta_-(\beta_+-\beta_-)} \right)}{\sqrt{R^2-\alpha \beta_-^2}} + \frac{ \mathrm{arctanh}\, \left (\frac{\sqrt{\alpha R^2 +\alpha^2 \beta_-^2}(\beta_--\beta_+)}{R^2+\alpha\beta_-(\beta_--\beta_+)} \right)}{\sqrt{R^2+\alpha \beta_-^2}}\right]\,.
\label{eq:pottanh}
\end{equation}
\end{widetext}

Given the highly challenging differential equation for $R(\tau)$ posed by \eqref{eq:pottanh}, we may only hope to obtain a numerical profile for $R(\tau)$. Nevertheless, we may indeed obtain an analytic expression for the potential $V(R)$ near infinity and for sufficiently small $R$. On the one hand, asymptotically the potential $V(R)$ takes the form
\begin{equation}
\begin{aligned}
V(R)&=-\frac{\mathsf{M}}{R^{(D-3)}}-\frac{\mathsf{m}^2}{4R^{2(D-3)}}\\
&+\frac{\alpha^2}{6 R^{4}}\left(\frac{\mathsf{m}^6}{10 R^{6(D-3)}}+\frac{\mathsf{m}^4 \mathsf{M}}{ R^{5(D-3)}}+\frac{18\mathsf{m}^2 \mathsf{M}^2}{5 R^{4(D-3)}} \right. \\& \qquad \qquad \left. +\frac{4\mathsf{M}^3}{R^{3(D-3)}}\right)+\mathcal{O}(\alpha^4)\, ,
\end{aligned}
\end{equation}
On the other hand, obtaining the potential for small $R$ turns out to be particularly difficult, as the function $\tanh \left( x^{-1} \right)$ is not analytic at $x=0$ (although $\lim_{x \rightarrow 0^+} \tanh \left( x^{-1} \right) =1$ and $\lim_{x \rightarrow 0^+} \frac{\mathrm{d}}{\mathrm{d}x^k} \left[\tanh \left( x^{-1} \right)\right] =0$ for all $k>0$). As a matter of fact, one may derive the following behavior of $V(R)$ near $R=0$:
\begin{equation}
\begin{aligned}
V(R)&=-\frac{R^2}{\alpha}+ \frac{\log(2)R^{D+1}}{2\alpha^2 \mathsf{M}} \\&+\mathcal{O} \left (R^{D+3}, e^{-4M \alpha/R^{D-1}} \right) \,.
\end{aligned}
\end{equation}
Hence, we obtain the very same leading quadratic piece that we found in the previous case, but the subleading terms are naturally different, and they contain non-analytic exponential terms. Interestingly, we find no evidence of $\log(R)$ terms this time, which implies that $V(R)$ is $\mathcal{C}^{\infty}$. 
The motion of a collapsing shell shares the same qualitative features as in the Hayward black hole case (observe the graphs in Fig.~\ref{fig:Rtau}), so we refer to the previous subsection for a detailed explanation of the dynamical collapse of the shell.

\subsubsection{General considerations}\label{gcons}
As one could suspect, some of the qualitative properties of the last two examples are generic, and do not depend on the particular theory and solution considered. The most relevant observation is that the behavior of $V(R)$ near $R=0$ is universal. 
To prove this, we recall from our earlier discussion around \req{R2psi0} that, in all theories with regular black holes, the solution behaves as 
\begin{equation}\label{fr0}
f(r)=1-\psi_0 r^2+\ldots
\end{equation}
around $r=0$, where $\psi_0$ is independent of the mass. 
Then, it is most useful to consider the form \req{sdd3} of the shell equation. 
For a given value of $\mathsf{M}$, if $R$ is taken sufficiently small, then \req{fr0} holds in most of the integration domain. Since $\psi_0$ is independent of the mass, the integral in \req{sdd3} can be carried out trivially, and we get
\begin{equation}
\mathsf{m}=\frac{\mathsf{M}}{\sqrt{\dot{R}^2+1-\psi_0 R^2}}+\ldots\, ,
\end{equation}
where the ellipsis denote subleading terms in the $R\to 0$ limit. Therefore, we conclude that
\begin{equation}
V(R)=-\psi_0 R^2+\ldots 
\end{equation}
This result is important, because it implies that the resolution of the singularity of thin-shell collapse is generic in these theories.  
Furthermore, since the potential vanishes at $R=0$ and at $R\to \infty$, this implies that the potential must necessarily have at least one minimum at some intermediate value of $R$.\footnote{Here we are naturally assuming that the solution for the potential exists for all values of $R$.} We suspect that, for all regular black holes with two horizons, the potential has exactly one minimum, and therefore always has the shape of the potentials in Fig.~\ref{fig:VR}. It would be interesting to find examples with additional minima, which would perhaps require additional horizons.  

Finally, it can also be observed that, if the shell experiences a bounce, $\dot{R}=0$, this must always happen in a ``static region'' with $f(R)>0$. This follows from the fact that the shell equation would become complex if $\dot{R}=0$ and $f(r)<0$; for instance, in \req{sdd}, $\beta_{+}$ would become imaginary.  In the case of the typical two-horizon regular black holes, this implies that the shell always bounces in the ``de Sitter core'' after crossing the inner horizon, and it never bounces in the intermediate region $r_{-}<r<r_{+}$.

\section{Causal structure}

Let us now discuss the causal structure of the regular spacetimes arising from the dynamical collapse of a thin shell in the infinite tower of Birkhoff theories given by \eqref{QTaction}. Such a casual structure will heavily depend on the number of event horizons displayed by the solution. For the sake of simplicity, we will restrict to those theories for which $f(r)$ has at most two zeros in the region $+\infty >r \geq 0$, as it is the case for all regular black holes considered in the manuscript. Within this set of theories, we will keep our discussion as theory-independent as possible, describing the main generic qualitative features of the causal structure of the subsequent regular spacetimes. Depending on the value of the mass of the solution, one may have non-extremal black holes, showcasing outer and inner horizons, extremal solutions, in which these horizons merge, or solitonic configurations. This latter case is somewhat trivial, as the causal structure would be equivalent to that of Minkowski spacetime. Let us then focus on the causal structure of non-extremal and extremal regular black holes arising from the collapse of a spherical thin shell. 

\subsection{Non-extremal black holes}

Let us consider the collapse of a thin shell into a non-extremal black hole with outer horizon at $r=r_+$ and inner horizon at $r=r_-$. To this aim, we may think of a sufficiently advanced civilization that compiles enough matter into a spherical thin shell, which is freely set to collapse at some moment. Let $R_0$ be the initial position of the shell. By gravitational collapse, the shell will shrink into smaller values of $r$, crossing at some point $r=r_+$ and forming the outer horizon. As the coordinates used in \eqref{eq:ssans} are singular at $r_+$, one would need to resort to Kruskal-Szekeres-like coordinates to describe the region around $r=r_+$. Nevertheless, differently from the situation with an eternal (regular) black hole, the introduction of these coordinates does not provide an extension of the original spacetime, since prior to the formation of the thin shell there was no black hole. However, the situation changes dramatically as the shell crosses $r=r_-$. From that moment on,  the inner horizon is formed and one would need to use a different set of Kruskal-Szekeres-like coordinates to patch a neighborhood of $r=r_-$. This new set of coordinates does allow one to extend the original spacetime, obtaining a new region which is causally disconnected from the interior of the thin shell and contains an $r=0$ hypersurface  (see region III' in Fig. \ref{fig:pds}, left). Although the interior of the shell is Minkowski spacetime, the new region III' corresponds to the interior of the inner horizon, with no singularity whatsoever --- as particles hit $r=0$ there, one can always continue worldlines through antipodal identification. In particular, the inner horizon is no longer a Cauchy horizon. Afterwards, the shell reaches a minimum radius $R_{\rm min}$, concludes its collapse and initiates a bounce towards the recovery of its initial size. Specifically, it crosses the other $r=r_-$ hypersurface created through the Kruskal-Szekeres extension and approaches a new $r=r_+$ horizon in the region II' (cf. Fig. \ref{fig:pds}, left). This corresponds to a white hole horizon and the shell emerges into a new universe. The shell will continue its growth up to $r=R_0$, from which the collapse and bounce of the shell is replicated \emph{ad infinitum}, leaving behind myriads of new universes.

\subsection{Extremal black holes}

\begin{figure}[t!]
 \centering
              \includegraphics[scale=0.47,trim={0 0.18cm 0 0},clip]{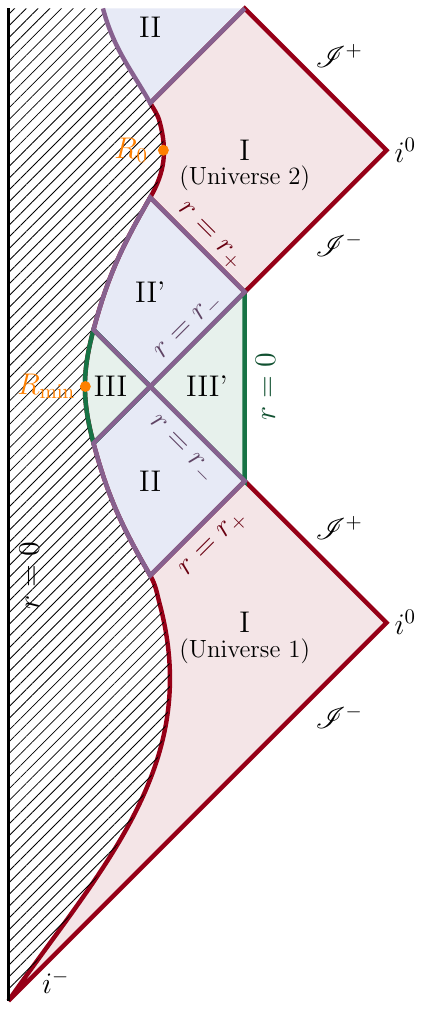}           
              \includegraphics[scale=0.47,trim={0 0.18cm 0 0},clip]{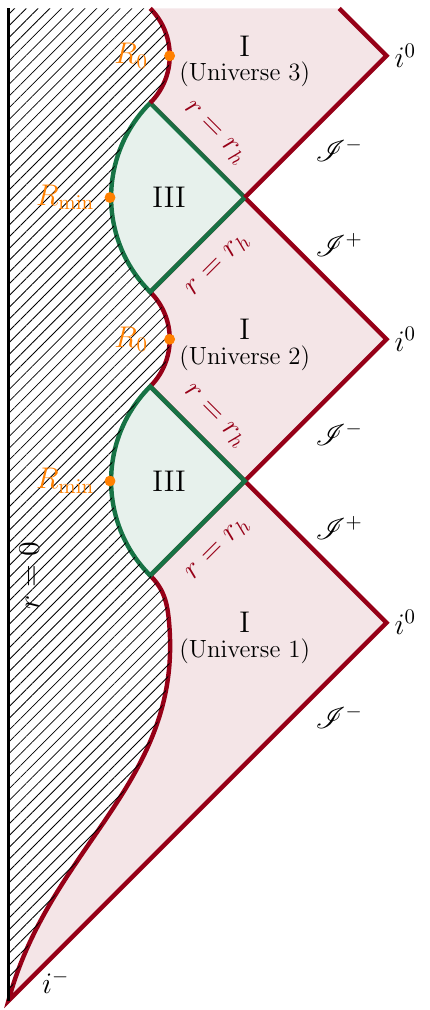}           
              \caption{Penrose diagrams associated with the dynamical collapse of a spherical thin shell in theories of the form \eqref{QTaction} whose spherically symmetric black holes display two event horizons. Left: non-extremal black hole. Right: extremal black hole. In both cases, matter is assumed to exist from past infinity and congregated into a spherical shell at some point, from which the shell initiates its collase. $R_0$ stands for the initial radius of the shell (corresponding to its maximum size), while $R_{\rm min}$ stands for the minimum radius, at which the shell initiates its bounce.}
              \label{fig:pds}
          \end{figure}

Let us now assume that the mass of the spherical thin shell is tuned to produce an extremal black hole, in which the outer and inner horizons merge into a single one, say at $r=r_h$. If the shell starts its collapse from rest at $r=R_0$, it will shrink and eventually approach and trespass $r=r_h$, giving rise to an extremal black hole. Afterwards, the shell will acquire its minimum size and begin its bounce back towards the initial radius. By extending the original spacetime with the aid of the Kruskal-Szekeres coordinates, the shell will approach a new $r=r_h$ hypersurface that corresponds to a while hole horizon. The shell will come out in a new universe, in which it will reattain its maximum size at $r=R_0$. The collapse is then restarted and the shell will undergo through an infinite sequence of collapses and bounces into new universes. This situation is depicted in the Penrose diagram shown in Fig. \ref{fig:pds} (right).

\section*{Discussion}

Exactly how --- or even \textit{if} --- the singularities of General Relativity are resolved in Nature remains a fundamental problem without a clear answer. It could be the case, for example, that singularity resolution is truly ``quantum'', occurring at a stage when a classical metric description is no longer valid. On the other hand, it could be the case that singularity resolution occurs when classical metrics and fields are still sensible concepts, at least approximately. In the absence of the ability to experimentally probe these questions, if either --- or another --- perspective is to be taken seriously, it becomes crucially important to construct explicit examples that demonstrate the physics at hand. It is this latter perspective to which our results apply. 

We have shown that the Schwarzschild singularity is \textit{generically} resolved when one incorporates into the action an infinite tower of higher curvature corrections, corrections general enough that they provide a basis for gravitational effective field theory in vacuum. More than this, we have demonstrated that the resulting black holes are the \textit{unique} black hole solutions of the corresponding theories, owing to a Birkhoff theorem. Most notably, we have shown these regular black holes are precisely those objects that are formed when matter collapses. This is the first example, to the best of our knowledge, of a model of such generality to appear in the literature, and as a consequence of well-motived purely gravitational theories. Our model emphatically shows that singularities \textit{can} be resolved, and without fine tuning or invoking new forms of matter. Altogether, this lends  credibility to the second option for singularity resolution mentioned above. 

The model and tools developed here offer considerable opportunity for further developments. A natural extension of this work would be to relax the assumption of a shell of dust, allowing for shells of matter with pressure. A further extension would be to consider the problem of multiple shells. Even in General Relativity, this simple extension of a one-body to two-body problem presents rich dynamics, including eternally oscillating shells or black hole formation exhibiting critical phenomena~\cite{Cardoso:2016wcr}. It would be interesting to understand the implications of higher curvature corrections and singularity resolution on these results. 

An important problem would be to assess the stability of these regular black holes from the higher-dimensional perspective. This is very challenging for several reasons. One must consider the entire tower of higher curvature corrections, as truncating the model at any finite order in curvature may introduce spurious instabilities. This is made more subtle due to the degeneracy properties of quasi-topological theories. As we explained earlier, at each order in curvature there can exist several distinct densities that give rise to the same equations of motion in spherical symmetry. This degeneracy is due to the existence of curvature invariants that vanish identically in spherical symmetry. Hence, adding to a given quasi-topological theory one of these “spherical-trivial” invariants results in a different covariant action, but does not alter the equations of motion in spherical symmetry, dimensional reduction on the sphere, or regular black hole solutions. However, these terms will contribute to the perturbations outside of the spherical sector. A thorough analysis would therefore require a complete classification of spherical-trivial densities, which would be an interesting mathematical problem.

Another chief aspect concerns inner horizons in regular black hole solutions. As is well known in the case of General Relativity, infalling matter is blueshifted to enormous energy densities at inner (Cauchy) horizons, the backreaction of which is expected to result in singularities (thereby enforcing strong cosmic censorship). Blueshift instabilities have a kinematical origin, but how the geometry reacts to the build up of energy is fundamentally a dynamical question. It is therefore not obvious that the same mass inflation phenomena that backreacts in a singular fashion in General Relativity will do so also in a theory capable of resolving curvature singularities. Our model can be used to assess this dynamical question, insofar as it can be addressed in a spherically symmetric setting. Of course, if the black hole forms dynamically and ultimately evaporates quantum mechanically, the inner horizon may not be as problematic as often thought.

A notable feature of all regular black holes is that they admit critical, or extremal, limits when the inner and event horizons coincide. These solutions bear many similarities to familiar extremal black holes of General Relativity, including having zero Hawking temperature, an AdS$_2$ throat, and (apparently) finite entropy. It has been recently understood that quantum gravitational fluctuations become large for extremal black holes in General Relativity, driving the entropy at extremality to zero in the absence of supersymmetry~\cite{Iliesiu:2020qvm}. It is natural to expect that quantum gravitational fluctuations will be important for these critical black holes as well. Using the two-dimensional dilaton gravity theory identified here, it would be possible to obtain a Jackiw-Teitelboim-like limit and understand the implications of strong quantum fluctuations on these black holes. These considerations would be especially relevant for understanding the final stages of the Hawking evaporation process for regular black holes, as Hawking radiation will gradually drive a non-extremal solution toward the extremal limit. It is known that, in these final moments, there can be large bursts of radiation~\cite{Frolov:2016gwl, Frolov:2017rjz} which these strong quantum gravitational fluctuations are perhaps sufficient to tame.

We have observed in a case-by-case analysis that in each of the resummations studied there is a universal, solution-independent upper bound on the Kretschmann scalar. This observation is suggestive of Markov's limiting curvature hypothesis. It would be interesting establish this connection more rigorously, or to find counterexamples. There are several possibilities for an exploration of this kind. For example, one may consider the restriction to the two-dimensional dilaton theory and attempt to prove a version of the limiting curvature hypothesis in this setting. A more ambitious goal would be to attempt to establish a similar bound in the resummed higher dimensional theory, considering, for example, geometries beyond spherical symmetry.

Our model requires very little in terms of constraints on the coupling constants of the higher curvature terms. Going forward, it would be interesting to use physical considerations to determine whether some resummations are more viable than others.  Along these lines, it is worth mentioning that not \textit{all} singularities in General Relativity are necessarily bad. For example, Horowitz and Myers argued that the resolution of the negative mass Schwarzschild singularity \textit{could} signal that the theory under consideration lacks a well-defined ground state~\cite{Horowitz:1995ta}. One could imagine constraining the viable resummations by the requirement that negative mass solutions remain singular. This is the case for the resummation resulting in the Hayward black hole, but certainly not the case for all possible choices.

It is simultaneously remarkable and puzzling that the same, purely gravitational mechanism can resolve the Schwarzschild singularity in all dimensions \textit{larger than four}.  From a physical perspective, it is perhaps unexpected that resolving singularities in higher dimensions is somehow `easier' than resolving singularities in four dimensions, as the gravitational potential becomes \textit{more} singular with increasing dimension. From a mathematical perspective, the fact that our mechanism does not apply in four dimensions is simply because quasi-topological theories do not exist in four dimensions and there are no theories that afford the same level of simplicity. It may be that this is only a technical issue that can be overcome by brute force calculation. But it is also conceivable that this is hinting at a deeper complexity of the singularities in a four-dimensional world.

\vspace{0.1cm}
\begin{acknowledgments} 
We would like to thank Javier Moreno, Simon Ross and Guido van der Velde for useful conversations and comments on this manuscript. PB was supported by a Ramón y Cajal fellowship (RYC2020-028756-I), by a Proyecto de Consolidación Investigadora (CNS 2023-143822) from Spain’s Ministry of Science, Innovation and Universities, and by the grant PID2022-136224NB-C22, funded by MCIN/AEI/ 10.13039/501100011033/FEDER, UE.
The work of PAC received the support of a fellowship from “la Caixa” Foundation (ID 100010434) with code LCF/BQ/PI23/11970032. 
The work of RAH received the support of a fellowship from ``la Caixa” Foundation (ID 100010434) and from the European Union’s Horizon 2020 research and innovation programme under the Marie Skłodowska-Curie grant agreement No 847648 under fellowship code LCF/BQ/PI21/11830027. \'AJM  was supported by a Juan de la Cierva contract (JDC2023-050770-I) from Spain’s Ministry of Science, Innovation and Universities. \'AJM would like to thank the University of Barcelona for its warm hospitality before the start of the contract.
\end{acknowledgments}

\onecolumngrid
\vspace{1cm}
\begin{center}  
{\Large\bf Appendices} 
\end{center} 
\appendix 

\section{Evaluation of higher-curvature invariants on a SS metric}\label{app2}
In this appendix we present the explicit expressions resulting from the evaluation of all invariants appearing in the Quasi-topological densities \eqref{Znexplicit} on a general $D$-dimensional spherically symmetric ansatz of the form \req{sphericmetric}. We find
\begin{subequations}\label{curvinv}
\begin{align}
W_{abcd} W^{abcd}&=\frac{(D-1)(D-2)^2(D-3) \Omega^2}{4}\,, \quad 
Z_{ab} Z^{ab} = (D-2) \Theta^2+\mathcal{S}_{\mu \nu} \mathcal{S}^{\mu \nu}\,,\\
W_{abcd} W^{cdef}W_{ef}{}^{ab}&=\frac{(D-1)(D-2)(D-3)(D^3-9 D^2+26D-22) \Omega^3}{8}\,, \\
W^{abcd} W_{ebcd} Z_a^e&=-\frac{(D-1)^2(D-2)(D-3)(D-4)\Theta \Omega^2}{8}\,, \\
W_{abcd} Z^{ac} Z^{bd}&=\frac{(D-2)(D-3)((D^2-2D+2)\Theta^2-\mathcal{S}_{\mu \nu} \mathcal{S}^{\mu \nu}) \Omega}{4}\,, \\
Z_{ac}Z^{cb} Z_b^a &=\frac{(D-2)((D^2-4D+6)\Theta^2-3\mathcal{S}_{\mu \nu} \mathcal{S}^{\mu \nu}) \Theta}{2}\,,\\
W_{acbd} W^{c efg} W^d{}_{efg} Z^{ab} &=-\frac{D(D-1)^2(D-2)(D-3)^2(D-4) \Theta \Omega^3}{32}\,, \\
W_{abcd} W^{aecf} Z^{bd} Z_{ef} &=\frac{(D-2)(D-3)^2((D-2)(3 D \Theta^2+\mathcal{S}_{\mu \nu} \mathcal{S}^{\mu \nu})+4 \Theta^2)\Omega^2}{16}\,,\\
Z_{a c} Z_{d e} W^{bdce} Z^{a}_b &=-\frac{(D-2)(D-3)(D-4)((D^2-2D+2) \Theta^2-\mathcal{S}_{\mu \nu} \mathcal{S}^{\mu \nu})\Theta \Omega}{8}\,,\\
Z_{a}^b Z_b^c Z_c^d Z_d^a&=(D-2) \Theta^4+\frac{(D-2)^4 \Theta^4+2(D-2)^2 \Theta^2\mathcal{S}_{\mu \nu} \mathcal{S}^{\mu \nu}+\left (\mathcal{S}_{\mu \nu} \mathcal{S}^{\mu \nu} \right)^2)}{2}\,,\\
Z^a_b Z^b_c W_{daef} W^{efgh} W_{gh}{}^{dc} &=\frac{(D^3-6D^2+11D-6)((D-3)^3 \mathcal{S}_{\mu \nu} \mathcal{S}^{\mu \nu}-2(D-5)\Theta^2)\Omega^3}{16}\,, \\
W_{a c d e}W^{bcde} Z^a_b Z_c^d Z_d^c&=-\frac{(D-2)(D-3)^2((D^4-4D^3+8D-8)\Theta^3-(D^2-4)\Theta \mathcal{S}_{\mu \nu} \mathcal{S}^{\mu \nu})\Omega^2}{32}\,, \\
Z_{a}^b Z_{b}^{c} Z_{cd} Z_{ef} W^{eafd}&=-\frac{(D-2)(D-3)((D-4)\Theta^2-\mathcal{S}_{\mu \nu} \mathcal{S}^{\mu \nu})((D^2-2D+2)\Theta^2-\mathcal{S}_{\mu \nu} \mathcal{S}^{\mu \nu})\Omega}{8}\,, \\
Z_{a}^b Z_b^c Z_c^d Z_d^e Z_e^a&=\frac{(D-2)\Theta( ( (D-2)^4+4)\Theta^4-5 \left ( \mathcal{S}_{\mu \nu} \mathcal{S}^{\mu \nu}\right)^2 )}{4}\,.
\end{align}
\end{subequations}
where the relevant objects appearing in the right-hand side of the equations are defined in section \ref{subi}.

\section{Regular black holes in even dimensions}\label{ee}
As explained in the main text, Birkhoff-QT gravities admit SSS black hole solutions characterized by a single function
\begin{equation}
\mathrm{d}s^2=-f(r)\mathrm{d}t^2+\frac{\mathrm{d}r^2}{f(r)}+r^2 \mathrm{d}\Omega_{(D-2)}^2\, ,
\end{equation}
where the metric $f(r)$ is determined by the algebraic equation
\begin{equation}\label{eomg}
h(\psi)=  \frac{ 2 {\mathsf M}}{r^{D-1}}\, , \quad \text{where}\quad h(\psi)\equiv \psi + \sum_{n=2}^{\infty} \alpha_n \frac{(D-2n)}{(D-2)}\psi^n\, , \quad \psi\equiv \frac{1-f(r)}{r^2}\, .
 \end{equation}
Convenient choices of $\alpha_n $ satisfying the generic properties explained above yield simple analytic expressions for $f(r)$\cite{Bueno:2024dgm}. However, a subtlety arises when $D$ is even. In that case, there exists a curvature order, corresponding to $n=D/2$, for which  QT gravities make no contribution to the characteristic polynomial $h(\psi)$ \cite{Bueno:2019ycr}. In other words --- and analogously to Lovelock theories ---
 QT gravities of order $n=D/2$ make no contribution to the equation of $f(r)$. While this does not affect the behavior near $r=0$ and therefore all the conclusions obtained in \cite{Bueno:2024dgm} regarding singularity resolution as well as analyticity properties of the different solutions remain valid, the functional form of $f(r)$ does get modified.
 
 \begin{figure*}
\centering \hspace{0cm}
\includegraphics[width=0.49\textwidth]{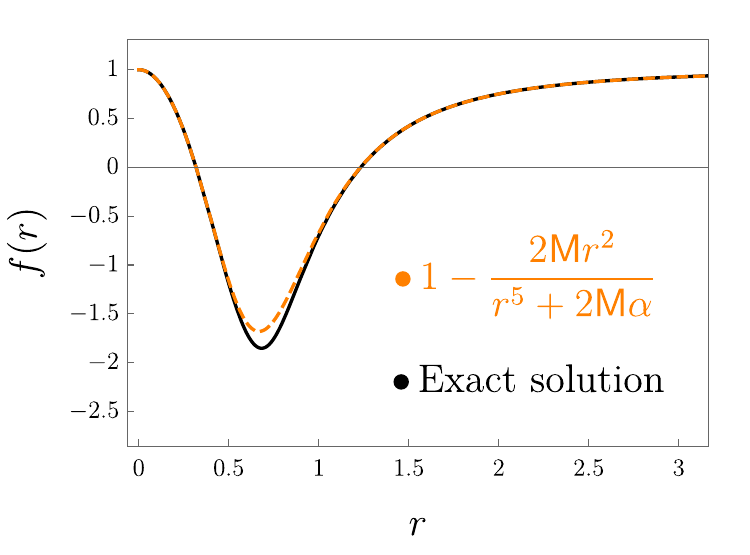}
\includegraphics[width=0.49\textwidth]{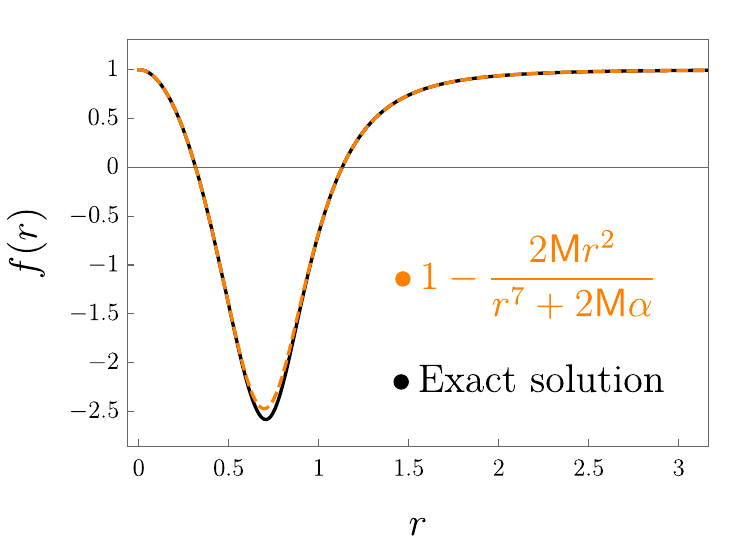}
\caption{We plot the metric function of regular black hole solutions to the Birkhoff-QT action \req{QTaction} for the choice of couplings in \req{oddd} in $D=6$ (Left) and $D=8$ (Right). In each case, the black curve corresponds to the exact solution, which was obtained by taking into account that the $n=D/2$ term is absent from $h(\psi)$. The orange dotted curves correspond to the extrapolation to those dimensions of the Hayward metric function which is an exact solution in odd dimensions (namely, the metric function which one obtains by artificially including the corresponding $n=D/2$ terms into $h(\psi)$). In all cases we set $ {\mathsf M}=1$, $\alpha=1/10$.   }
\label{fig:aaay}
\end{figure*}

 Consider for instance the case of a Hayward black hole. This is achieved, for odd $D$, by choosing 
 \begin{equation}\label{oddd}
 \alpha_n=\frac{(D-2)}{(D-2n)}\alpha^{n-1} \quad \Rightarrow \quad  h(\psi)=\sum_{n=1}^{\infty}\alpha^{n-1}\psi^n=\frac{\psi}{1-\alpha \psi}\quad \Rightarrow \quad f(r)=1- \frac{ 2 {\mathsf M}r^2}{r^{D-1}+2 {\mathsf M} \alpha} \quad ({\rm odd}\,\, D)\end{equation}
On the other hand, the corresponding characteristic polynomial in the even $D$ case reads instead
\begin{equation}\label{evend}
h(\psi)= \sum_{n=1}^{D/2-1}\alpha^{n-1}\psi^n + \sum_{n=D/2+1}^{\infty}\alpha^{n-1}\psi^n=\frac{\psi}{1-\alpha \psi}-\alpha^{\frac{D-2}{2}}\psi^{\frac{D}{2}}\, , \quad ({\rm even}\,\, D)
\end{equation}
namely, one needs to remove the $n=D/2$ term from the sum. As anticipated, this new form of $h(\psi)$ does not allow \req{eomg} to be solved analytically for $f(r)$ for general values of $D$. In Fig.\,\ref{fig:aaay} we show the results obtained in $D=6,8$ from solving for $f(r)$ the even-dimensional equation with the actual characteristic polynomial given by \req{evend} and how they compare with the extrapolation obtained from the simpler odd-dimensional case of \req{oddd}. As expected, the Hayward-like approximation is excellent both asymptotically and near $r=0$, but becomes worse in the intermediate region, where the effects of the missing $n=D/2$ term are more relevant. In both cases it is possible to find the exact solutions analytically, but the form of $f(r)$ is way messier than the Hayward one. 

Simpler explicit solutions can be found in even dimensions by making appropriate choices of the couplings which avoid the problematic  $n=D/2$ term. Examples of this kind have been presented in the main text in \req{fN} and \req{tanhmodel}.


\bibliographystyle{JHEP-2}
\bibliography{Gravities}
\noindent 

\appendix
\onecolumngrid \vspace{1.5cm}

\end{document}
%